\pdfoutput=1
\documentclass[12pt,a4paper]{article}
\usepackage{ifthen} % for conditional statements
\newboolean{pdflatex}
\setboolean{pdflatex}{true} % False for eps figures 
\newboolean{articletitles}
\setboolean{articletitles}{true} % False removes titles in references
\newboolean{uprightparticles}
\setboolean{uprightparticles}{false} %True for upright particle symbols
\newboolean{inbibliography}
\setboolean{inbibliography}{false} %True once you enter the bibliography
\usepackage[top=1in, bottom=1.25in, left=1in, right=1in]{geometry}

\usepackage{microtype}
\usepackage{lineno}  % for line numbering during review
\usepackage{xspace} % To avoid problems with missing or double spaces after
\usepackage{caption} %these three command get the figure and table captions automatically small

\usepackage{graphicx}  % to include figures (can also use other packages)
\usepackage{color}
\usepackage{colortbl}
\graphicspath{{./figs/}} % Make Latex search fig subdir for figures

\usepackage{amsmath} % Adds a large collection of math symbols
\usepackage{amssymb}
\usepackage{amsfonts}
\usepackage{upgreek} % Adds in support for greek letters in roman typeset

\newcommand*\patchAmsMathEnvironmentForLineno[1]{%
\expandafter\let\csname old#1\expandafter\endcsname\csname #1\endcsname
\expandafter\let\csname oldend#1\expandafter\endcsname\csname
end#1\endcsname
 \renewenvironment{#1}%
   {\linenomath\csname old#1\endcsname}%
   {\csname oldend#1\endcsname\endlinenomath}%
}
\newcommand*\patchBothAmsMathEnvironmentsForLineno[1]{%
  \patchAmsMathEnvironmentForLineno{#1}%
  \patchAmsMathEnvironmentForLineno{#1*}%
}
\AtBeginDocument{%
\patchBothAmsMathEnvironmentsForLineno{equation}%
\patchBothAmsMathEnvironmentsForLineno{align}%
\patchBothAmsMathEnvironmentsForLineno{flalign}%
\patchBothAmsMathEnvironmentsForLineno{alignat}%
\patchBothAmsMathEnvironmentsForLineno{gather}%
\patchBothAmsMathEnvironmentsForLineno{multline}%
\patchBothAmsMathEnvironmentsForLineno{eqnarray}%
}

\usepackage{hyperref}    % Hyperlinks in references
\usepackage[all]{hypcap} % Internal hyperlinks to floats.

\def\mpmm       {\ensuremath{\Pmu^+ \Pmu^-}\xspace}
\def\lplm       {\ensuremath{\ell^+ \ell^-}\xspace}
\def\lp       {\ensuremath{\ell^+ }\xspace}
\def\ln       {\ensuremath{\ell^- }\xspace}

\def\RKst    {\mbox{R_{\Kstarz}}}
\def\RK    {\mbox{R_{K}}}
\def\lhcb {\mbox{LHCb}\xspace}

\ifthenelse{\boolean{uprightparticles}}%
{

 \def\Pmu         {\ensuremath{\upmu}\xspace}

 \def\Ppi         {\ensuremath{\uppi}\xspace}

 \def\Ppsi        {\ensuremath{\uppsi}\xspace}

 \def\PDelta      {\ensuremath{\Delta}\xspace}                 
 \def\PXi      {\ensuremath{\Xi}\xspace}                 
 \def\PLambda      {\ensuremath{\Lambda}\xspace}                 
 \def\PSigma      {\ensuremath{\Sigma}\xspace}                 
 \def\POmega      {\ensuremath{\Omega}\xspace}                 
 \def\PUpsilon      {\ensuremath{\Upsilon}\xspace}                 
 
 \def\PB      {\ensuremath{\mathrm{B}}\xspace}                 
                  
 \def\PD      {\ensuremath{\mathrm{D}}\xspace}

 \def\PJ      {\ensuremath{\mathrm{J}}\xspace}                 
 \def\PK      {\ensuremath{\mathrm{K}}\xspace}

 \def\PZ      {\ensuremath{\mathrm{Z}}\xspace}                 
                  
 \def\Pb      {\ensuremath{\mathrm{b}}\xspace}                 
 \def\Pc      {\ensuremath{\mathrm{c}}\xspace}                 
                  
 \def\Pe      {\ensuremath{\mathrm{e}}\xspace}

 \def\Pi      {\ensuremath{\mathrm{i}}\xspace}

 \def\Pp      {\ensuremath{\mathrm{p}}\xspace}

 \def\Ps      {\ensuremath{\mathrm{s}}\xspace}

}
{

 \def\Pmu         {\ensuremath{\mu}\xspace}

 \def\Ppi         {\ensuremath{\pi}\xspace}

 \def\Ppsi        {\ensuremath{\psi}\xspace}                 
                  
 \mathchardef\PDelta="7101
 \mathchardef\PXi="7104
 \mathchardef\PLambda="7103
 \mathchardef\PSigma="7106
 \mathchardef\POmega="710A
 \mathchardef\PUpsilon="7107
                  
 \def\PB      {\ensuremath{B}\xspace}                 
                  
 \def\PD      {\ensuremath{D}\xspace}

 \def\PJ      {\ensuremath{J}\xspace}                 
 \def\PK      {\ensuremath{K}\xspace}

 \def\PZ      {\ensuremath{Z}\xspace}                 
                  
 \def\Pb      {\ensuremath{b}\xspace}                 
 \def\Pc      {\ensuremath{c}\xspace}                 
                  
 \def\Pe      {\ensuremath{e}\xspace}

 \def\Pi      {\ensuremath{i}\xspace}

 \def\Pp      {\ensuremath{p}\xspace}

 \def\Ps      {\ensuremath{s}\xspace}

}

\def\en         {\ensuremath{\Pe^-}\xspace}   % electron negative (\em is taken)
\def\ep         {\ensuremath{\Pe^+}\xspace}
 
\def\epem       {\ensuremath{\Pe^+\Pe^-}\xspace}
\def\mumu       {\ensuremath{\Pmu^+\Pmu^-}\xspace}

\def\lepton     {\ensuremath{\ell}\xspace}
\def\ellm       {\ensuremath{\ell^-}\xspace}
\def\ellp       {\ensuremath{\ell^+}\xspace}
\def\Z      {\ensuremath{\PZ}\xspace}

\def\squark    {\ensuremath{\Ps}\xspace}

\def\cquark    {\ensuremath{\Pc}\xspace}

\def\bquark    {\ensuremath{\Pb}\xspace}

\def\pion  {\ensuremath{\Ppi}\xspace}

\def\pip   {\ensuremath{\pion^+}\xspace}
\def\pim   {\ensuremath{\pion^-}\xspace}

\def\kaon  {\ensuremath{\PK}\xspace}
  \def\Kbar  {\kern 0.2em\overline{\kern -0.2em \PK}{}\xspace}

\def\Kp    {\ensuremath{\kaon^+}\xspace}
\def\Km    {\ensuremath{\kaon^-}\xspace}

\def\Kstarz  {\ensuremath{\kaon^{*0}}\xspace}

\def\Kstar   {\ensuremath{\kaon^*}\xspace}

  \def\Dbar    {\kern 0.2em\overline{\kern -0.2em \PD}{}\xspace}
\def\D       {\ensuremath{\PD}\xspace}

\def\Dz      {\ensuremath{\D^0}\xspace}

\def\Dm      {\ensuremath{\D^-}\xspace}

\def\Dstarp  {\ensuremath{\D^{*+}}\xspace}

\def\B       {\ensuremath{\PB}\xspace}
\def\Bbar    {\ensuremath{\kern 0.18em\overline{\kern -0.18em \PB}{}}\xspace}

\def\Bz      {\ensuremath{\B^0}\xspace}

\def\Bu      {\ensuremath{\B^+}\xspace}

\def\Bd      {\ensuremath{\B^0}\xspace}
\def\Bs      {\ensuremath{\B^0_\squark}\xspace}
\def\Bsb     {\ensuremath{\Bbar^0_\squark}\xspace}

\def\jpsi     {\ensuremath{{\PJ\mskip -3mu/\mskip -2mu\Ppsi\mskip 2mu}}\xspace}
\def\psitwos  {\ensuremath{\Ppsi{(2S)}}\xspace}

  \def\Y#1S{\ensuremath{\PUpsilon{(#1S)}}\xspace}% no space before {...}!

\def\proton      {\ensuremath{\Pp}\xspace}
\def\antiproton  {\ensuremath{\overline \proton}\xspace}

\def\Lbar {\ensuremath{\kern 0.1em\overline{\kern -0.1em\PLambda}}\xspace}

\def\Lbbar   {{\ensuremath{\Lbar{}^0_\bquark}}\xspace}

\def\BF         {{\ensuremath{\cal B}\xspace}}

\def\BR         {\BF}
\newcommand{\decay}[2]{\ensuremath{#1\!\to #2}\xspace}         % {\Pa}{\Pb \Pc}

\def\to                 {\ensuremath{\rightarrow}\xspace}

\def\qsq       {\ensuremath{q^2}\xspace}

\def\BdToKstmm    {\decay{\Bd}{\Kstarz\mup\mun}}

\def\AT#1     {\ensuremath{A_{\mathrm{T}}^{#1}}\xspace}           % 2

\def\ctl       {\ensuremath{\cos{\theta_\ell}}\xspace}

\def\C#1      {\ensuremath{\mathcal{C}_{#1}}\xspace}                       % 9
\def\Cp#1     {\ensuremath{\mathcal{C}_{#1}^{'}}\xspace}                    % 7
\def\Ceff#1   {\ensuremath{\mathcal{C}_{#1}^{\mathrm{(eff)}}}\xspace}        % 9  
\def\Cpeff#1  {\ensuremath{\mathcal{C}_{#1}^{'\mathrm{(eff)}}}\xspace}       % 7
\def\Ope#1    {\ensuremath{\mathcal{O}_{#1}}\xspace}                       % 2
\def\Opep#1   {\ensuremath{\mathcal{O}_{#1}^{'}}\xspace}                    % 7

\newcommand{\tev}{\ifthenelse{\boolean{inbibliography}}{\ensuremath{~T\kern -0.05em eV}\xspace}{\ensuremath{\mathrm{\,Te\kern -0.1em V}}\xspace}}
\newcommand{\gev}{\ensuremath{\mathrm{\,Ge\kern -0.1em V}}\xspace}
\newcommand{\mev}{\ensuremath{\mathrm{\,Me\kern -0.1em V}}\xspace}
\newcommand{\kev}{\ensuremath{\mathrm{\,ke\kern -0.1em V}}\xspace}
\newcommand{\ev}{\ensuremath{\mathrm{\,e\kern -0.1em V}}\xspace}
\newcommand{\gevc}{\ensuremath{{\mathrm{\,Ge\kern -0.1em V\!/}c}}\xspace}
\newcommand{\mevc}{\ensuremath{{\mathrm{\,Me\kern -0.1em V\!/}c}}\xspace}
\newcommand{\gevcc}{\ensuremath{{\mathrm{\,Ge\kern -0.1em V\!/}c^2}}\xspace}
\newcommand{\gevgevcccc}{\ensuremath{{\mathrm{\,Ge\kern -0.1em V^2\!/}c^4}}\xspace}
\newcommand{\mevcc}{\ensuremath{{\mathrm{\,Me\kern -0.1em V\!/}c^2}}\xspace}
\def\mum  {\ensuremath{{\,\upmu\rm m}}\xspace}

\def\invfb   {\ensuremath{\mbox{\,fb}^{-1}}\xspace}

\newcommand{\stat}{\ensuremath{\mathrm{\,(stat)}}\xspace}
\newcommand{\syst}{\ensuremath{\mathrm{\,(syst)}}\xspace}

\newcommand{\chisq}{\ensuremath{\chi^2}\xspace}

\newcommand{\chisqip}{\ensuremath{\chi^2_{\rm IP}}\xspace}

\def\gsim{{~\raise.15em\hbox{$>$}\kern-.85em
          \lower.35em\hbox{$\sim$}~}\xspace}
\def\lsim{{~\raise.15em\hbox{$<$}\kern-.85em
          \lower.35em\hbox{$\sim$}~}\xspace}

\def\sPlot{\mbox{\em sPlot}}

\def\ptot       {\mbox{$p$}\xspace}
\def\pt         {\mbox{$p_{\rm T}$}\xspace}
\def\et         {\mbox{$E_{\rm T}$}\xspace}

\def\evtgen     {\mbox{\textsc{EvtGen}}\xspace}

\def\geant      {\mbox{\textsc{Geant4}}\xspace}

\def\photos     {\mbox{\textsc{Photos}}\xspace}

\def\pythia     {\mbox{\textsc{Pythia}}\xspace}

\def\tell1  {TELL1\xspace}
\def\ukl1   {UKL1\xspace}

\usepackage{cite} % Allows for ranges in citations
\usepackage{mciteplus}

\def\lqsq{low-\qsq}
\def\cqsq{central-\qsq}

\def\loe{\textrm{L0E}\xspace}
\def\loh{\textrm{L0H}\xspace}
\def\loi{\textrm{L0I}\xspace}

\def\RJPs{\ensuremath{r_{\jpsi}}\xspace}
\def\RCC{\ensuremath{R_{\psitwos}}\xspace}

\def\RH{\ensuremath{R_{H}}\xspace}
\def\RKst{\ensuremath{R_{\Kstarz}}\xspace}
\def\RK{\ensuremath{R_{\kaon}}\xspace}
\def\RKK{\ensuremath{R_{\kaon^{(*)}}}\xspace}

\def\pp{\ensuremath{\proton\proton}\xspace}

\def\mKpi{\ensuremath{m(\KPi)}\xspace}

\def\mKpill{\ensuremath{m(\KPi\ellell)}\xspace}
\def\mKpimm{\ensuremath{m(\KPi \mumu)}\xspace}
\def\mKpiee{\ensuremath{m(\KPi \ee)}\xspace}

\def\mcorr{\ensuremath{m_{\textrm{corr}}}\xspace}
\def\chisqfd{\ensuremath{\chi^2_{\rm VD}}\xspace}

\def\KPi{\ensuremath{\Kp\pim}\xspace}
\def\KPill{\ensuremath{\KPi\ellell}\xspace}

\def\KPiee{\ensuremath{\KPi\ee}\xspace}

\def\ll{\ensuremath{\ellp\ellm}\xspace}
\def\ellell{\ensuremath{\ellp\ellm}\xspace}
 
\def\ee{\ensuremath{\ep\en}\xspace}

\def\JPsll{\ensuremath{\jpsi(\ellell)}\xspace}

\def\Kone{\ensuremath{\kaon_{1}^{+}(1270)}\xspace} 
\def\Ktwo{\ensuremath{\kaon_{2}^{*+}(1430)}\xspace} 

\def\BdToKstVll{\mbox{\decay{\Bd}{\Kstarz V(\decay{}{\ll})}}\xspace}

\def\BdToKstll{\mbox{\decay{\Bd}{\Kstarz \ll}}\xspace}
\def\BdToKstmm{\mbox{\decay{\Bd}{\Kstarz \mumu}}\xspace}
\def\BdToKstee{\mbox{\decay{\Bd}{\Kstarz \epem}}\xspace}

\def\BdToKstG{\mbox{\decay{\Bd}{\Kstarz \gamma}}\xspace}
\def\BdToKstGee{\mbox{\decay{\Bd}{\Kstarz \gamma(\to\epem)}}\xspace}

\def\BdToKstJPsll{\mbox{\decay{\Bd}{\Kstarz \jpsi(\decay{}{\ll})}}\xspace}
\def\BdToKstJPsmm{\mbox{\decay{\Bd}{\Kstarz \jpsi(\decay{}{\mumu})}}\xspace}
\def\BdToKstJPsee{\mbox{\decay{\Bd}{\Kstarz \jpsi(\to\epem)}}\xspace}

\def\BToXmm{\mbox{\decay{\B}{\Kstarz \mumu X}}\xspace}
\def\BToXee{\mbox{\decay{\B}{X(\to Y \Kstarz)\epem}}\xspace}

\def\BdToKstPsill{\mbox{\decay{\Bd}{\Kstarz \psitwos(\to\ll)}}\xspace}
\def\BdToKstPsimm{\mbox{\decay{\Bd}{\Kstarz \psitwos(\to\mumu)}}\xspace}

\def\BuToKll{\mbox{\decay{\Bu}{\Kp \ll}}\xspace}

\def\BuToKJPsll{\mbox{\decay{\Bu}{\Kp \jpsi(\decay{}{\ll})}}\xspace}

\def\BsToPhill{\mbox{\decay{\Bs}{\phi\ll}}\xspace}

\def\BsToKstJPsmm{\mbox{\decay{\Bsb}{\Kstarz \jpsi(\decay{}{\mumu})}}\xspace}
\def\BsToKstJPsee{\mbox{\decay{\Bsb}{\Kstarz \jpsi(\decay{}{\ee})}}\xspace}

\def\LbTopKJPsmm{\mbox{\decay{\Lbbar}{\Kp\antiproton \jpsi(\decay{}{\mumu})}}\xspace}
\def\LbTopKJPsee{\mbox{\decay{\Lbbar}{\Kp\antiproton \jpsi(\to\epem)}}\xspace}

\usepackage{bigints}
\usepackage{multirow}
\begin{document}
%%%%%%%%%%%%%%%%%%%%%%%%%
\renewcommand{\thefootnote}{\fnsymbol{footnote}}
\setcounter{footnote}{1}
%%%%%%%%%%%%%%%%%%%%%%%%%%%%%%%%%%%%
% !TEX root = main.tex
%%%%%%%%%%%%%%%%%%%%%%%%%%%%%%%%%%%%

\begin{titlepage}
\pagenumbering{roman}

\vspace*{-1.5cm}
\centerline{\large EUROPEAN ORGANIZATION FOR NUCLEAR RESEARCH (CERN)}
\vspace*{1.5cm}
\noindent
\begin{tabular*}{\linewidth}{lc@{\extracolsep{\fill}}r@{\extracolsep{0pt}}}
\ifthenelse{\boolean{pdflatex}}
{\vspace*{-2.7cm}\mbox{\!\!\!\includegraphics[width=.14\textwidth]{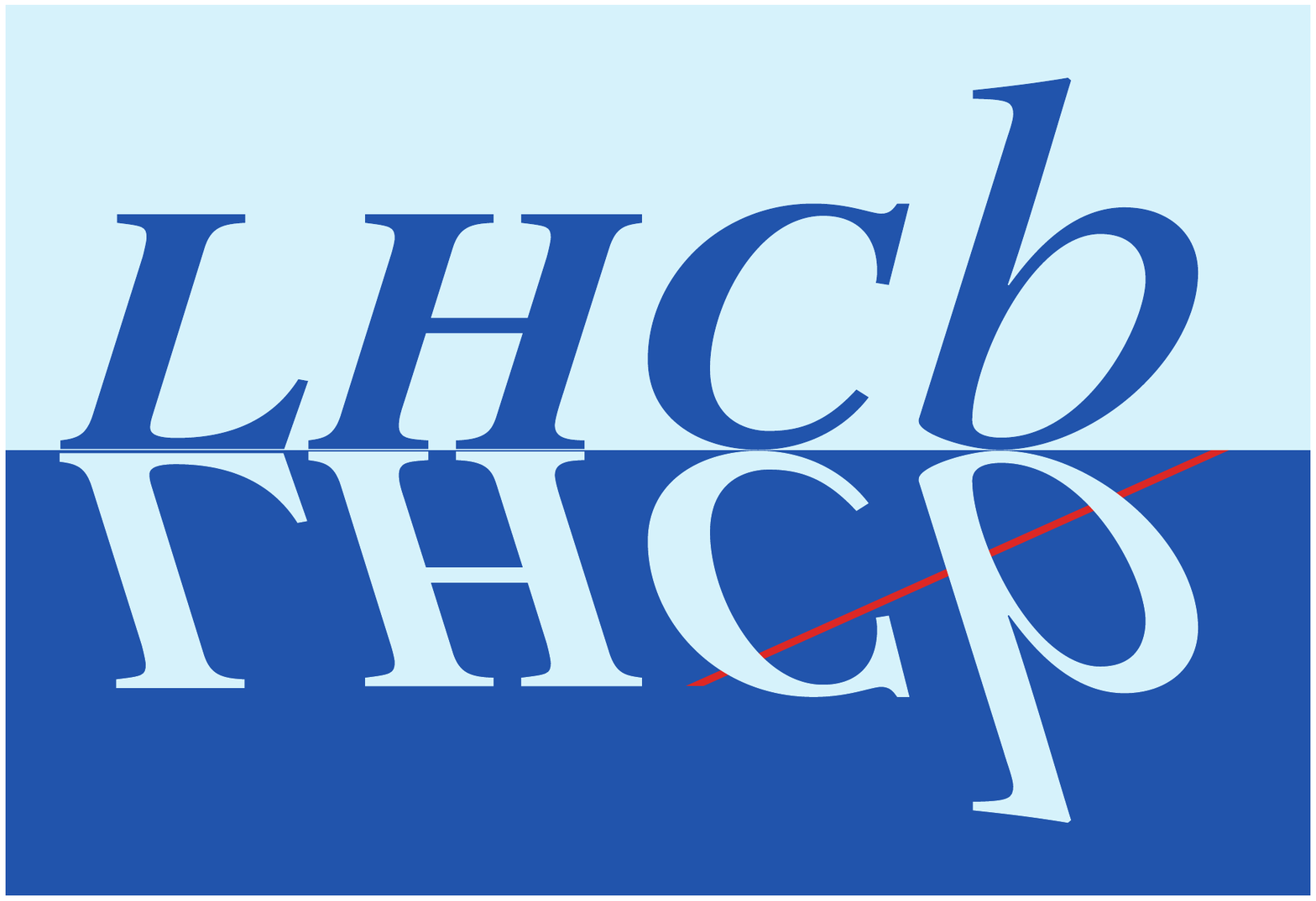}} & &}%
{\vspace*{-1.2cm}\mbox{\!\!\!\includegraphics[width=.12\textwidth]{lhcb-logo.eps}} & &}%
\\
 & & CERN-EP-2017-100 \\ 
 & & LHCb-PAPER-2017-013 \\
% & & May 16, 2017 \\
 & & August 22, 2017 \\
 & & \\
\end{tabular*}

\vspace*{2.0cm}

{\normalfont\bfseries\boldmath\huge
\begin{center}
Test of lepton universality with $B^{0} \rightarrow K^{*0}\ell^{+}\ell^{-}$ decays
\end{center}
}

\vspace*{1.cm}

\begin{center}
The LHCb collaboration\footnote{Authors are listed at the end of this paper.}
\end{center}

\vspace{\fill}

\begin{abstract}
\noindent
A test of lepton universality, performed by measuring the ratio of the branching fractions of the $B^{0} \rightarrow K^{*0}\mu^{+}\mu^{-}$ and $B^{0} \rightarrow K^{*0}e^{+}e^{-}$ decays, $R_{K^{*0}}$, is presented.
The $K^{*0}$ meson is reconstructed in the final state $K^{+}\pi^{-}$, which is required to have an invariant mass within 100$\mathrm{\,Me\kern -0.1em V\!/}c^2$ of the known $K^{*}(892)^{0}$ mass. 
The analysis is performed using proton-proton collision data, corresponding to an integrated luminosity of about 3$\mathrm{\,fb}^{-1}$, collected by the LHCb experiment at centre-of-mass energies of 7 and 8$\mathrm{\,Te\kern -0.1em V}$. 
The ratio is measured in two regions of the dilepton invariant mass squared, $q^{2}$, to be
\begin{eqnarray*}
R_{K^{*0}} = 
\begin{cases}
0.66~^{+~0.11}_{-~0.07}\mathrm{\,(stat)}  \pm 0.03\mathrm{\,(syst)}	& \textrm{for } 0.045 < q^{2} < 1.1~\mathrm{\,Ge\kern -0.1em V^2\!/}c^4 \, , \\
0.69~^{+~0.11}_{-~0.07}\mathrm{\,(stat)} \pm 0.05\mathrm{\,(syst)}	& \textrm{for } 1.1\phantom{00} < q^{2} < 6.0~\mathrm{\,Ge\kern -0.1em V^2\!/}c^4 \, .
\end{cases}
\end{eqnarray*}
The corresponding 95.4\% confidence level intervals are $[0.52, 0.89]$ and $[0.53, 0.94]$.
The results, which represent the most precise measurements of $R_{K^{*0}}$ to date, are compatible with the Standard Model expectations at the level of 2.1--2.3 and 2.4--2.5 standard deviations in the two $q^{2}$ regions, respectively.
\end{abstract}

\vspace*{1.5cm}

\begin{center}
Published in JHEP 08 (2017) 055
\end{center}

\vspace{\fill}

{\footnotesize 
\centerline{\copyright~CERN on behalf of the \lhcb collaboration, licence \href{http://creativecommons.org/licenses/by/4.0/}{CC-BY-4.0}.}}
\vspace*{2mm}

\end{titlepage}

\newpage
\setcounter{page}{2}
\mbox{~}

\cleardoublepage

\renewcommand{\thefootnote}{\arabic{footnote}}
\setcounter{footnote}{0}
%%%%%%%%%%%%%%%%%%%%%%%%%
\pagestyle{plain}
\setcounter{page}{1}
\pagenumbering{arabic}
%\linenumbers
%%%%%%%%%%%%%%%%%%%%%%%%%
%%%%%%%%%%%%%%%%%%%%%%%%%%%%%%%%%%%%
% !TEX root = main.tex
%%%%%%%%%%%%%%%%%%%%%%%%%%%%%%%%%%%%

\section{Introduction}

In the Standard Model (SM) of particle physics, the electroweak couplings of leptons to gauge bosons are independent of their flavour and the model is referred to as exhibiting lepton universality (LU).
Flavour-changing neutral-current (FCNC) processes, where a quark changes its flavour without altering its electric charge, provide an ideal laboratory to test LU.
The SM forbids FCNCs at tree level and only allows amplitudes involving electroweak loop (penguin and box) Feynman diagrams.
The absence of a dominant tree-level SM contribution implies that such transitions are rare, and therefore sensitive to the existence of new particles.
The presence of such particles could lead to a sizeable increase or decrease in the rate of particular decays, or change the angular distribution of the final-state particles.
Particularly sensitive probes for such effects are ratios of the type~\cite{Hiller:2003js}
\begin{eqnarray*}
\RH = \frac{\bigintssss \frac{ d\Gamma(B \to H \mumu) }{d\qsq} \, d\qsq}{\bigintssss \frac{ d\Gamma(B \to H\epem) }{d\qsq} \, d\qsq} \, ,
\end{eqnarray*}
where $H$ represents a hadron containing an \squark quark, such as a \kaon or a \Kstar meson.  
The decay rate, $\Gamma$, is integrated over a range of the squared dilepton invariant mass, \qsq. 
The \RH ratios allow very precise tests of LU, as hadronic uncertainties in the theoretical predictions cancel, and are expected to be close to unity in the SM~\cite{Hiller:2003js,Bobeth:2007dw,Bouchard:2013mia}.

At \epem colliders operating at the $\PUpsilon(4S)$ resonance, the ratios \RKK have been measured   to be consistent with unity with a precision of 20 to 50\%~\cite{Lees:2012tva,Wei:2009zv}.  
More recently, the most precise determination to date of \RK in the \qsq range between 1.0 and 6.0\gevgevcccc has been performed by the \lhcb collaboration.
The measurement has a relative precision of 12\%~\cite{LHCb-PAPER-2014-024} and is found to be 2.6 standard deviations lower than the SM expectation~\cite{Hiller:2003js}.
Hints of LU violation have been observed in \mbox{$\decay{\B}{\D^{(*)}\ell\nu_{\ell}}$} decays~\cite{Aubert:2008yv,Huschle:2015rga,LHCB-PAPER-2015-025}.
Tensions with the SM have also been found in several measurements of branching fractions~\cite{LHCb-PAPER-2014-006, LHCb-PAPER-2015-023, LHCb-PAPER-2015-009} and angular observables~\cite{LHCb-PAPER-2015-051, Wehle:2016yoi} of rare \decay{\bquark}{\squark} decays.  
Models containing a new, neutral, heavy gauge boson~\cite{Descotes-Genon:2013wba,Gauld,Buras,Altmannshofer,Altmannshofer:2014cfa,Altmannshofer:2016jzy} or leptoquarks~\cite{Becirevic:2016yqi,Crivellin:2017zlb} have been proposed to explain these measurements.

\begin{figure}[t!]
\centering
\includegraphics[width=0.46\textwidth]{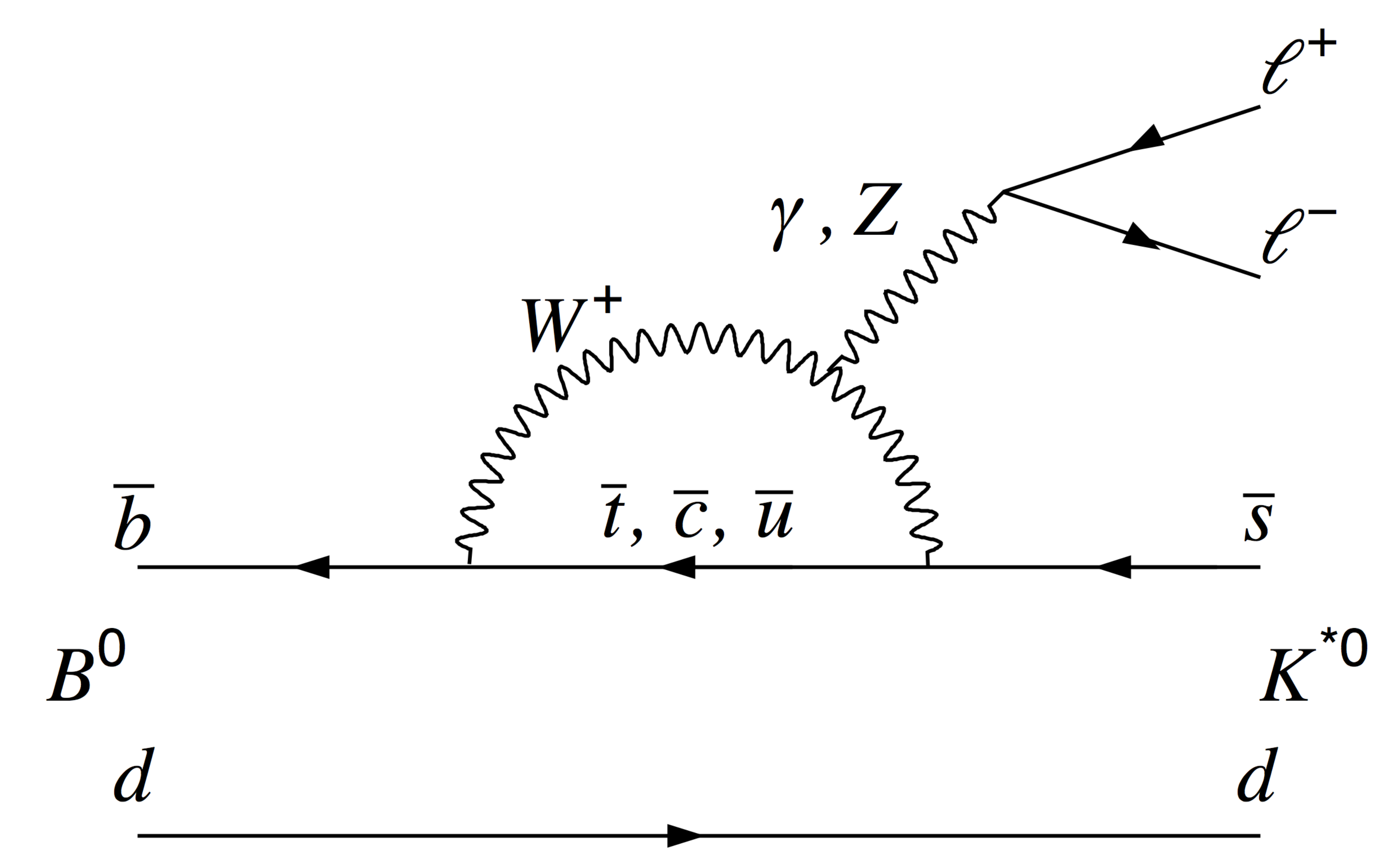} 
\includegraphics[width=0.46\textwidth]{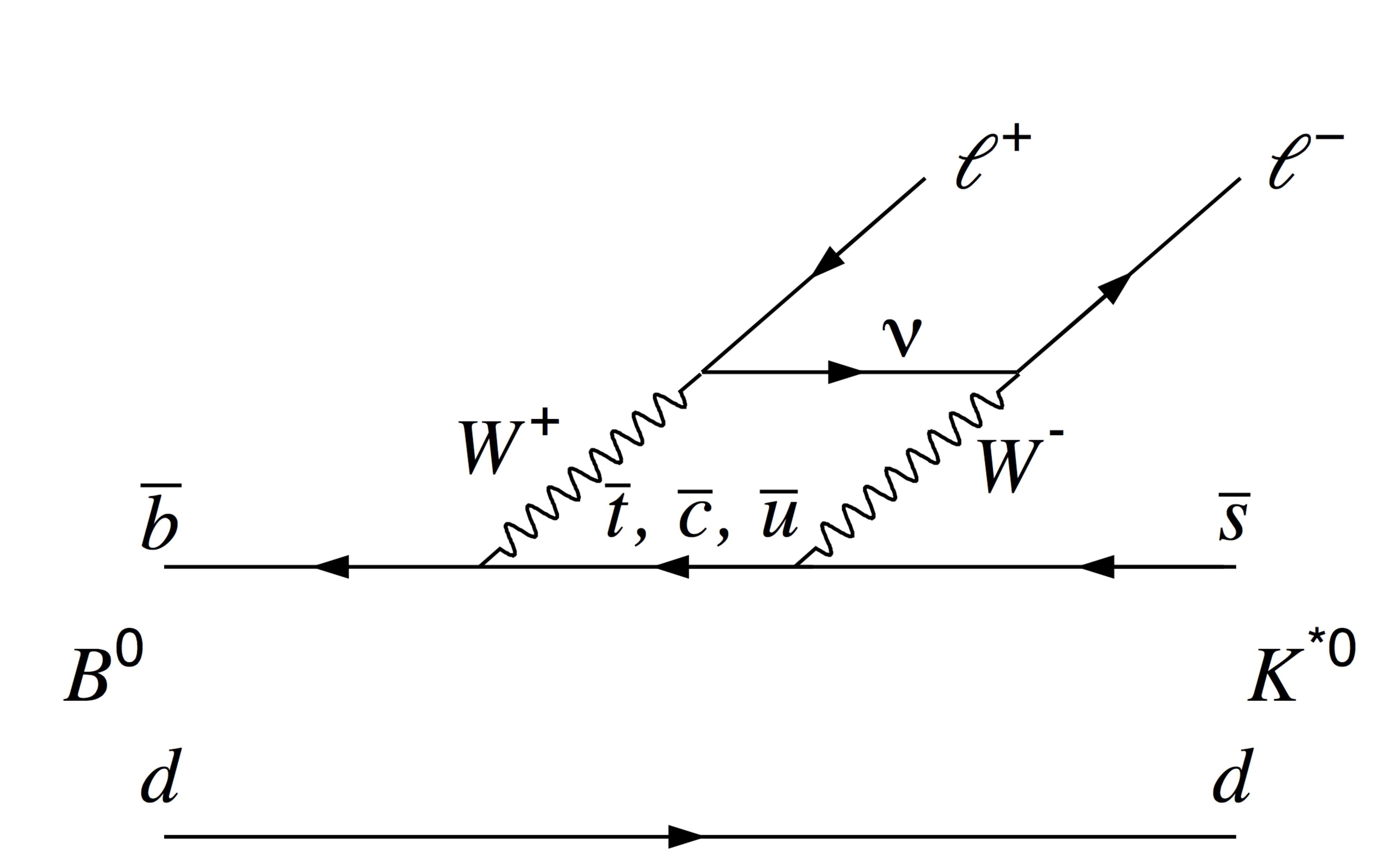}
\includegraphics[width=0.46\textwidth]{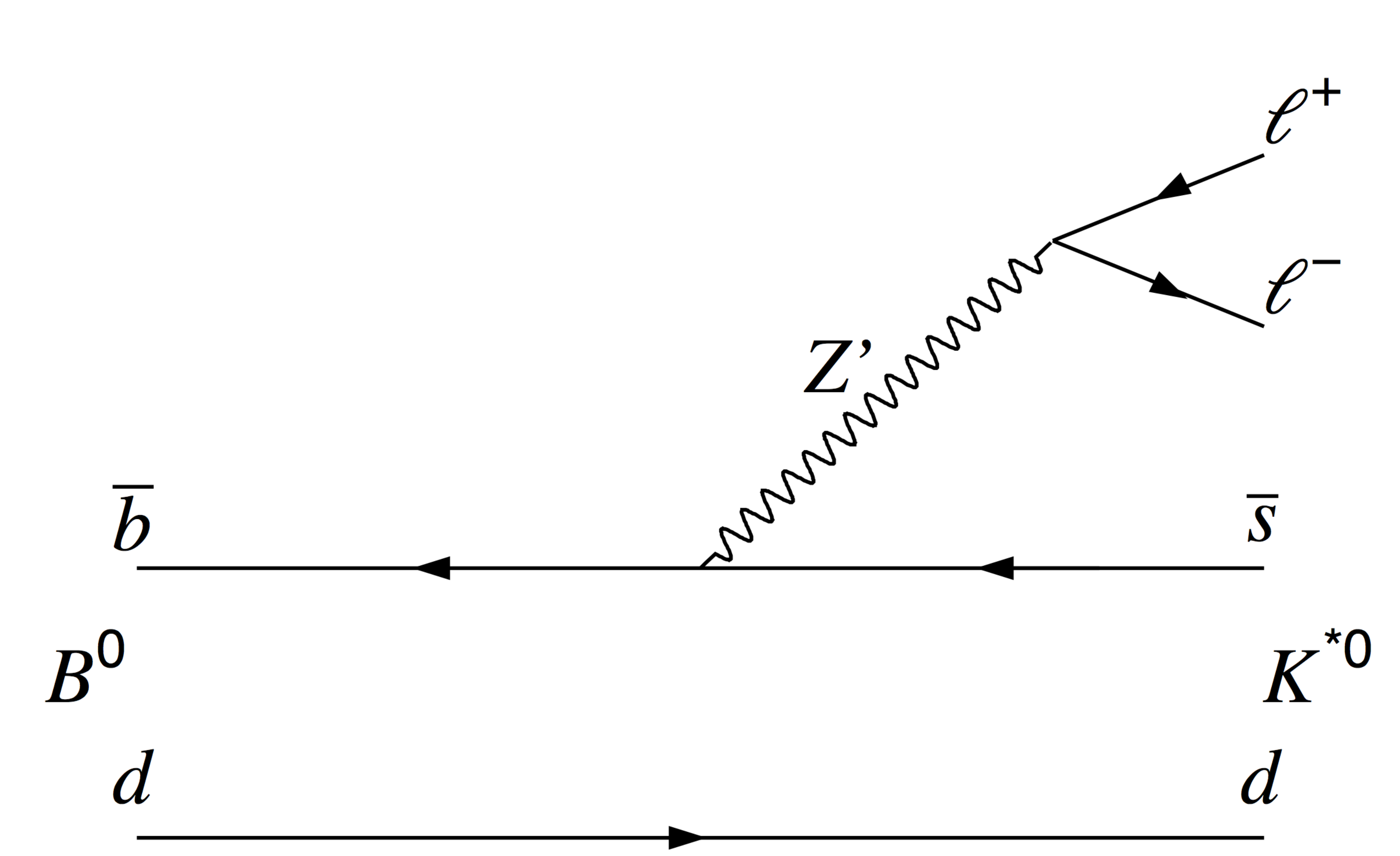}
\includegraphics[width=0.46\textwidth]{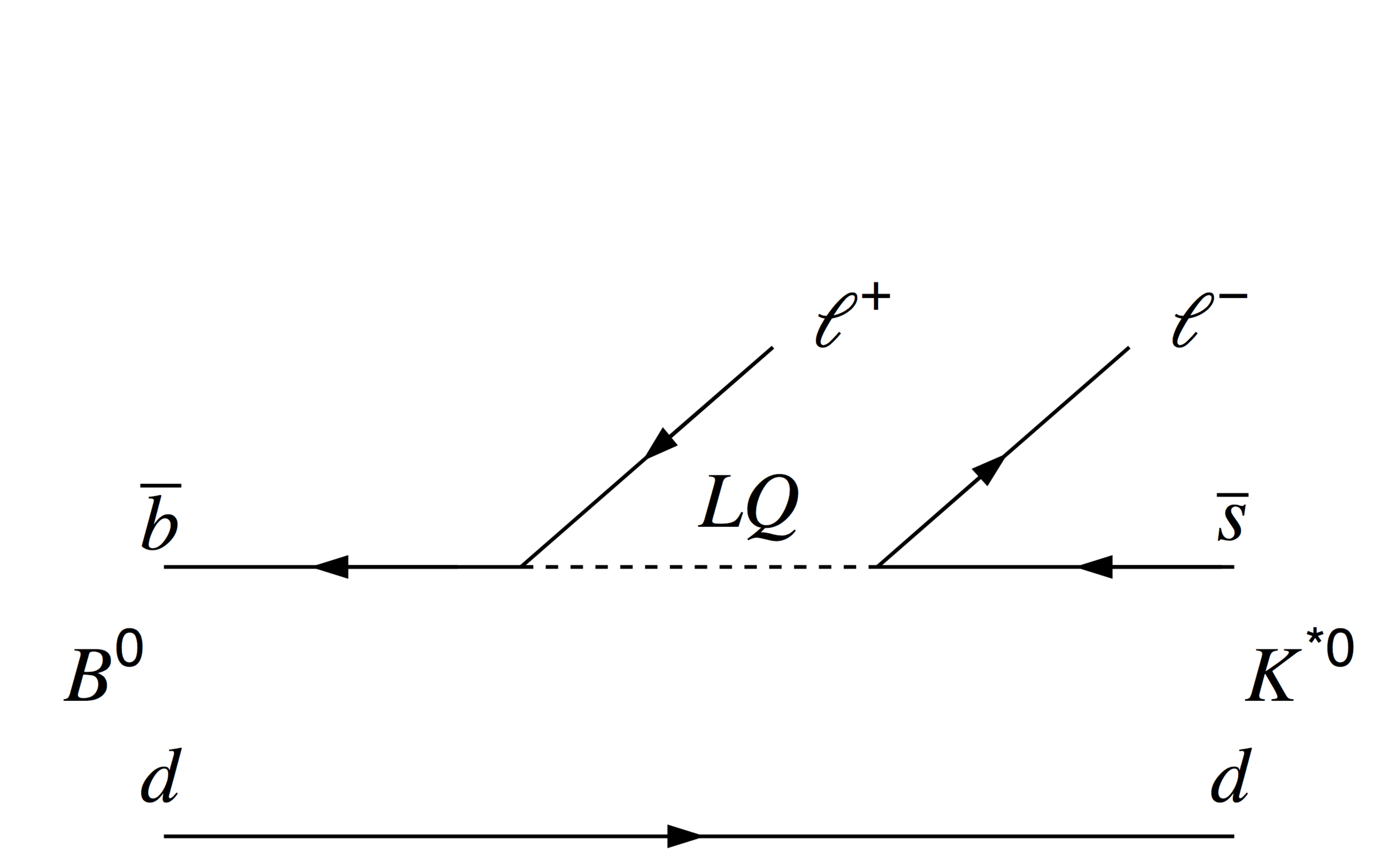}
\vspace{0.1cm}
\caption{Feynman diagrams in the SM of the \BdToKstll decay for the (top left) electroweak penguin and (top right) box diagram. Possible NP contributions violating LU: (bottom left) a tree-level diagram mediated by a new gauge boson $\Z^{\prime}$ and (bottom right) a tree-level diagram involving a leptoquark $LQ$.}
\label{fig:Feynman}
\end{figure}

A precise measurement of \RKst can provide a deeper understanding of the nature of the present discrepancies~\cite{Hiller:2014ula}.
Some of the leading-order Feynman diagrams for the \BdToKstll decays, where \lepton represents either a muon or an electron, are shown in figure~\ref{fig:Feynman} for both SM and possible New Physics (NP) scenarios.
If the NP particles couple differently to electrons and muons, LU could be violated.
The \Kstarz represents a $\Kstar(892)^{0}$ meson, which is reconstructed in the \KPi final state by selecting candidates within 100\mevcc of the known mass~\cite{Olive:2016xmw}.
No attempt is made to separate the \Kstarz meson from S-wave or other broad contributions present in the selected \KPi region.
The S-wave fraction contribution to the \BdToKstmm mode has been measured by the \lhcb collaboration and found to be small~\cite{LHCb-PAPER-2016-012}.
Inclusion of charge-conjugate processes is implied throughout the paper, unless stated otherwise.
The analysis is performed in two regions of \qsq that are sensitive to different NP contributions: a \lqsq bin, between 0.045 and 1.1\gevgevcccc, and a \cqsq bin, between 1.1 and 6.0\gevgevcccc.  
The lower boundary of the \lqsq region corresponds roughly to the dimuon kinematic threshold.
The boundary at 1.1\gevgevcccc is chosen such that \decay{\phi(1020)}{\ll} decays, which could potentially dilute NP effects, are included in the \lqsq interval.
The upper boundary of the \cqsq bin at 6.0\gevgevcccc is chosen to reduce contamination  from the radiative tail of the \jpsi resonance. 

The measurement is performed as a double ratio of the branching fractions of the \BdToKstll and \BdToKstJPsll decays
\begin{eqnarray*}
\RKst = {\frac{\BR(\BdToKstmm)}{\BR(\BdToKstJPsmm)}} \bigg{/} {\frac{\BR(\BdToKstee)}{\BR(\BdToKstJPsee)}} \, ,
\end{eqnarray*}
where the two channels are also referred to as the ``nonresonant'' and the ``resonant'' modes, respectively.
The experimental quantities relevant for the measurement are the yields and the reconstruction efficiencies of the four decays entering in the double ratio.
Due to the similarity between the experimental efficiencies of the nonresonant and resonant decay modes, many sources of systematic uncertainty are substantially reduced.
This helps to mitigate the significant differences in reconstruction between decays with muons or electrons in the final state, mostly due to bremsstrahlung emission and  the trigger response.
The decay \mbox{\decay{\jpsi}{\ll}} is measured to be consistent with LU~\cite{Olive:2016xmw}.
In order to avoid experimental biases, a blind analysis was performed.
The measurement is corrected for final-state radiation (FSR).
Recent SM predictions for \RKst in the two \qsq regions are reported in table~\ref{tab:predictions}.
Note that possible uncertainties related to QED corrections are only included in Ref.~\cite{Bordone:2016gaq}, and these are found to be at the percent level.
The \RKst ratio is smaller than unity in the \lqsq region due to phase-space effects.

The remainder of this paper is organised as follows:
section~\ref{sec:Detector} describes the \lhcb detector, as well as the data and the simulation samples used;
the experimental challenges in studying electrons as compared to muons are discussed in section~\ref{sec:eleRecoEff};
section~\ref{sec:weights} details how the simulation is adjusted in order to improve the modelling of the data;
the selection of the candidates, rejection of the background and extraction of the yields are outlined in sections~\ref{sec:selection}, \ref{sec:backgrounds} and \ref{sec:yields};
section~\ref{sec:eff} discusses the efficiency determination;
the cross-checks performed and the systematic uncertainties associated with the measurement are summarised in sections~\ref{sec:crosschecks} and \ref{sec:systematics}, respectively;
the results are presented in section~\ref{sec:results};
and section~\ref{sec:conclusions} presents the conclusions of the paper.

\begin{table}[t!]
\centering
\caption{Recent SM predictions for \RKst.}
\label{tab:predictions}
\renewcommand\arraystretch{1.4}
\begin{tabular}{c|l@{}c@{}l|r@{}l}
\qsq range $[\gevgevcccc\,]$ & \multicolumn{3}{c|}{$\RKst^{\textrm{\textbf{SM}}}$} & \multicolumn{2}{c}{References} \\
\hline
\multirow{5}{*}{$[0.045, 1.1]$}
& $0.906$				& $\pm$		& $\phantom{0}0.028$			& BIP & \cite{Bordone:2016gaq} \\
& $0.922$				& $\pm$		& $\phantom{0}0.022$			& CDHMV & \cite{Descotes-Genon:2015uva,Capdevila:2016ivx,Capdevila:2017ert} \\
& $0.919$				& $^{+}_{-}$	& $\phantom{0}^{0.004}_{0.003}$	& \texttt{EOS} & \cite{Serra:2016ivr,EOS-Web,*EOS} \\
& $0.925\phantom{00}$	& $\pm$		& $\phantom{0}0.004$			& \texttt{flav.io} & \cite{Straub:2015ica,Altmannshofer:2017fio,flavio} \\
& $0.920$				& $^{+}_{-}$	& $\phantom{0}^{0.007}_{0.006}$	& JC & \cite{Jager:2014rwa} \\
\hline
\multirow{5}{*}{$[1.1, 6.0]$}
& $1.000$				& $\pm$		& $\phantom{0}0.010$			& BIP & \cite{Bordone:2016gaq} \\
& $1.000$				& $\pm$		& $\phantom{0}0.006$			& CDHMV & \cite{Descotes-Genon:2015uva,Capdevila:2016ivx,Capdevila:2017ert} \\
& $0.9968$			& $^{+}_{-}$	& $\phantom{0}^{0.0005}_{0.0004}$	& \texttt{EOS} & \cite{Serra:2016ivr,EOS-Web,*EOS} \\
& $0.9964$			& $\pm$		& $\phantom{0}0.005$			& \texttt{flav.io} & \cite{Straub:2015ica,Altmannshofer:2017fio,flavio} \\
& $0.996$				& $\pm$		& $\phantom{0}{0.002}$			& JC & \cite{Jager:2014rwa} \\
\end{tabular}
\end{table}

%%%%%%%%%%%%%%%%%%%%%%%%%%%%%%%%%%%%
% !TEX root = main.tex
%%%%%%%%%%%%%%%%%%%%%%%%%%%%%%%%%%%%

\section{The \lhcb detector and data set}
\label{sec:Detector}

The \lhcb detector~\cite{Alves:2008zz,LHCb-DP-2014-002} is a single-arm forward
spectrometer covering the pseudorapidity range $2<\eta <5$, designed to study particles containing \bquark or \cquark quarks.
The detector includes a high-precision tracking system consisting of a silicon-strip vertex detector surrounding the \pp interaction region, a large-area silicon-strip detector located upstream of a dipole magnet with a bending power of about $4{\rm\,Tm}$, and three stations of silicon-strip detectors and straw
drift tubes placed downstream of the magnet.
The tracking system provides a measurement of momentum, \ptot, with a relative uncertainty that varies from 0.5\% at low values to 1.0\% at 200\gevc.
The minimum distance of a track to a primary vertex (PV), the impact parameter (IP), is measured with a resolution of $(15+29/\pt)\mum$, where \pt is the component of the momentum transverse to the beam, in \gevc.
Different types of charged hadrons are distinguished using information from two ring-imaging Cherenkov detectors.
Photons, electrons and hadrons are identified by a calorimeter system consisting of
scintillating-pad and preshower detectors, an electromagnetic calorimeter (ECAL) and a hadronic calorimeter (HCAL).
Muons are identified by a system composed of alternating layers of iron and multiwire
proportional chambers.

The trigger system consists of a hardware stage, based on information from the calorimeter and muon systems, followed by a software stage, which applies a full event reconstruction.
The hardware muon trigger selects events containing at least one muon with significant \pt (from $\sim1.5$ to $\sim1.8$\gevc, depending on the data-taking period).
The hardware electron trigger requires the presence of a cluster of calorimeter cells with significant transverse energy, \et, (from $\sim2.5$ to $\sim3.0$\gev, depending on the data-taking period) in the ECAL.
The hardware hadron trigger requires the presence of an energy deposit with \et above $\sim3.5$\gev in the calorimeters.
The software trigger requires a two-, three- or four-track secondary vertex, with a significant displacement from the PV.
At least one charged particle must have significant \pt and be inconsistent with originating from any PV.
A multivariate algorithm~\cite{BBDT} is used for the identification of secondary vertices consistent with the decay of a \bquark hadron.

The analysis is based on \pp collision data collected with the \lhcb detector at centre-of-mass energies of 7 and 8 \tev during 2011 and 2012,  and corresponding to an integrated luminosity of about 3\invfb.
Samples of simulated \BdToKstmm, \BdToKstee, \BdToKstJPsmm and \BdToKstJPsee events are used to determine the efficiency to trigger, reconstruct and select signal events, as well as to model the shapes used in the fits for signal candidates.
In addition, specific simulated samples are utilised to estimate the contributions from backgrounds and to model their mass distributions. 
The \pp collisions are generated using \pythia~\cite{Sjostrand:2006za,*Sjostrand:2007gs} with a specific \lhcb configuration~\cite{LHCb-PROC-2010-056}.
Decays of hadronic particles are described by \evtgen~\cite{Lange:2001uf}, in which FSR is generated using \photos~\cite{Golonka:2005pn}, which is observed to agree with a full QED calculation at the level of $\sim1\%$~\cite{Bordone:2016gaq}.
The interaction of the generated particles with the detector, and its response, are implemented using the \geant toolkit~\cite{Allison:2006ve, *Agostinelli:2002hh} as described in Ref.~\cite{LHCb-PROC-2011-006}.

%%%%%%%%%%%%%%%%%%%%%%%%%%%%%%%%%%%%
% !TEX root = main.tex
%%%%%%%%%%%%%%%%%%%%%%%%%%%%%%%%%%%%

\section{Electron reconstruction effects}
\label{sec:eleRecoEff}

The experimental environment in which the \lhcb detector operates leads to significant differences in the treatment of decays involving muons or electrons in the final state.
The two types of leptons behave differently when travelling through the detector material. 
Electrons emit a much larger amount of bremsstrahlung which, if not accounted for, would result in a significant degradation of the momentum resolution and consequently in a degradation of the \B mass resolution.
If the radiation occurs downstream of the dipole magnet, the photon energy is deposited in the same calorimeter cell as that of the lepton, and the momentum of the electron is correctly measured.
If the photons are emitted upstream of the magnet, the electron and photon deposit their energy in different calorimeter cells, and the electron momentum is evaluated after bremsstrahlung emission.
However, for both types of emissions, the ratio of the energy detected in the ECAL to the momentum measured by the tracking system, an important variable to identify electrons, remains unbiased.

A dedicated bremsstrahlung recovery procedure is used to improve the electron momentum reconstruction.
Searches are made within a region of the ECAL defined by the extrapolation of the electron track upstream of the magnet for energy deposits with $\et > 75\mev$ that are not associated with charged tracks.
Such ``bremsstrahlung clusters'' are added to the measured electron momentum.
If the same cluster can be associated with both the \ep and the \en, its energy is added to one of the two electrons at random.
In \BdToKstJPsee decays, one bremsstrahlung cluster is added to either electron of the pair in about half of the cases; the remaining half is equally split between cases when no bremsstrahlung cluster is found, or two or more clusters are added.
These fractions are reproduced well by the simulation and depend only weakly on \qsq. 
The bremsstrahlung recovery procedure is limited in three ways: the energy threshold of the clusters that are added; the calorimeter acceptance and resolution; and the presence of energy deposits wrongly interpreted as bremsstrahlung clusters.
These limitations degrade the resolution of the reconstructed invariant masses of both the dielectron pair and the \B candidate. 

Since the occupancy of the calorimeters is significantly higher than that of the muon stations, the constraints on the trigger rate require that higher thresholds are imposed on the electron \et than on the muon \pt.
In the \cqsq region the higher threshold causes a loss of about half of the electron signal.
The efficiency decreases slightly at lower \qsq values.
To partially mitigate this effect, decays with electrons in the final state can also be selected through the hadron hardware trigger, using clusters associated with the \Kstarz decay products, or by any hardware trigger from particles in the event that are not associated with the signal candidate.

In decays with electrons, since the mass resolution of the reconstructed \B candidate is worse than in final states with muons, the background contamination in the signal region is larger.
The level of combinatorial background, arising from the accidental association of particles produced by different \bquark- and \cquark-hadron decays, is also higher in such channels, due to a larger number of electron candidates.
As a result, the discriminating power of the fits to extract the signal yields is reduced (see section~\ref{sec:yields}).
Differences due to bremsstrahlung and the trigger response lead to a reconstruction efficiency for the \BdToKstJPsee decays that is about five times smaller than for the \BdToKstJPsmm decays.

\section{Corrections to the simulation}
\label{sec:weights}

In order to optimise the selection criteria and accurately evaluate the efficiencies, a set of corrections is determined from unbiased control samples selected from the data.
The procedure is applied to the simulated samples of the nonresonant and resonant modes. 

The first correction accounts for differences between simulation and data in the particle identification (PID) performance~\cite{LHCb-DP-2012-003}.
The PID efficiencies are directly measured using a tag-and-probe method on high-purity data samples of pions and kaons from \mbox{\decay{\Dstarp}{\Dz(\rightarrow \Km \pip)\pip}} decays.
Similarly, the electron and muon identification efficiencies are obtained from \BuToKJPsll decays.
Corrections are determined as a function of the track momentum and pseudorapidity. 

The second step of the procedure adjusts the simulation for the charged-track multiplicity in the event, which is not described well in simulation.
A small correction for the \Bz kinematics is also applied.
Resonant \BdToKstJPsmm decays are used since the muon triggers are observed to be well modelled in simulation.

The third step corrects the simulation of the trigger response for both the hardware and software levels using a tag-and-probe technique.
Whenever possible, \BdToKstJPsmm decays are used as a control sample in place of \BdToKstJPsee decays in order to take advantage of the larger sample size.
In such cases, the two decays are compared and found to give consistent results.
The tag sample is defined by events where the hardware trigger is fired by activity in the event not associated with any of the signal decay particles.
Alternatively, when probing the leptonic (hadronic) hardware triggers, the tag is required to have triggered the hadronic (leptonic) hardware trigger.
The corrections for the leptonic hardware triggers are parameterised as a function of the cluster \et or track \pt.
The hadron hardware trigger efficiency is known to be sensitive to tracks overlapping in the HCAL, however, a good description can be obtained when the efficiency is measured as a function of the \pt of the \KPi pair instead of the kaon or the pion independently.
Corrections are determined separately in the different calorimeter regions~\cite{Alves:2008zz}, in order to take into account potential differences due to different occupancies.
When the hardware trigger is fired by activity in the event not associated with any of the signal decay particles, the correction is determined as a function of the \Bz \pt and the charged-track multiplicity in the event in order to take into account correlations in the production between the two \bquark hadrons in the event.
For the software trigger, the corrections are determined as a function of the minimum \pt of the \Bz decay products. 

Finally, residual differences between data and simulation in the reconstruction performance are accounted for using \BdToKstJPsll candidates to which the full selection is applied, as well as additional requirements to further reduce the background contamination. 
The corrections are determined by matching the distribution of the \Bz kinematics and vertex fit quality in simulation to the data, separately for muon and electron samples. 

The correction factors are determined sequentially as histograms, with the previous corrections applied before deriving the subsequent one.
To avoid biases in the procedure due to common candidates being used for both the determination of the corrections and the measurement, a $k$-folding~\cite{Blum:1999:BHB:307400.307439} approach with $k = 10$ is adopted.
To dilute the dependence on the choice of the binning schemes, all corrections are linearly interpolated between adjacent bins.
After all the corrections are applied to the simulation, a very good agreement with the data is obtained.

\section{Selection of signal candidates}
\label{sec:selection}

\begin{figure}[t!]
\centering
\includegraphics[width=0.49\textwidth]{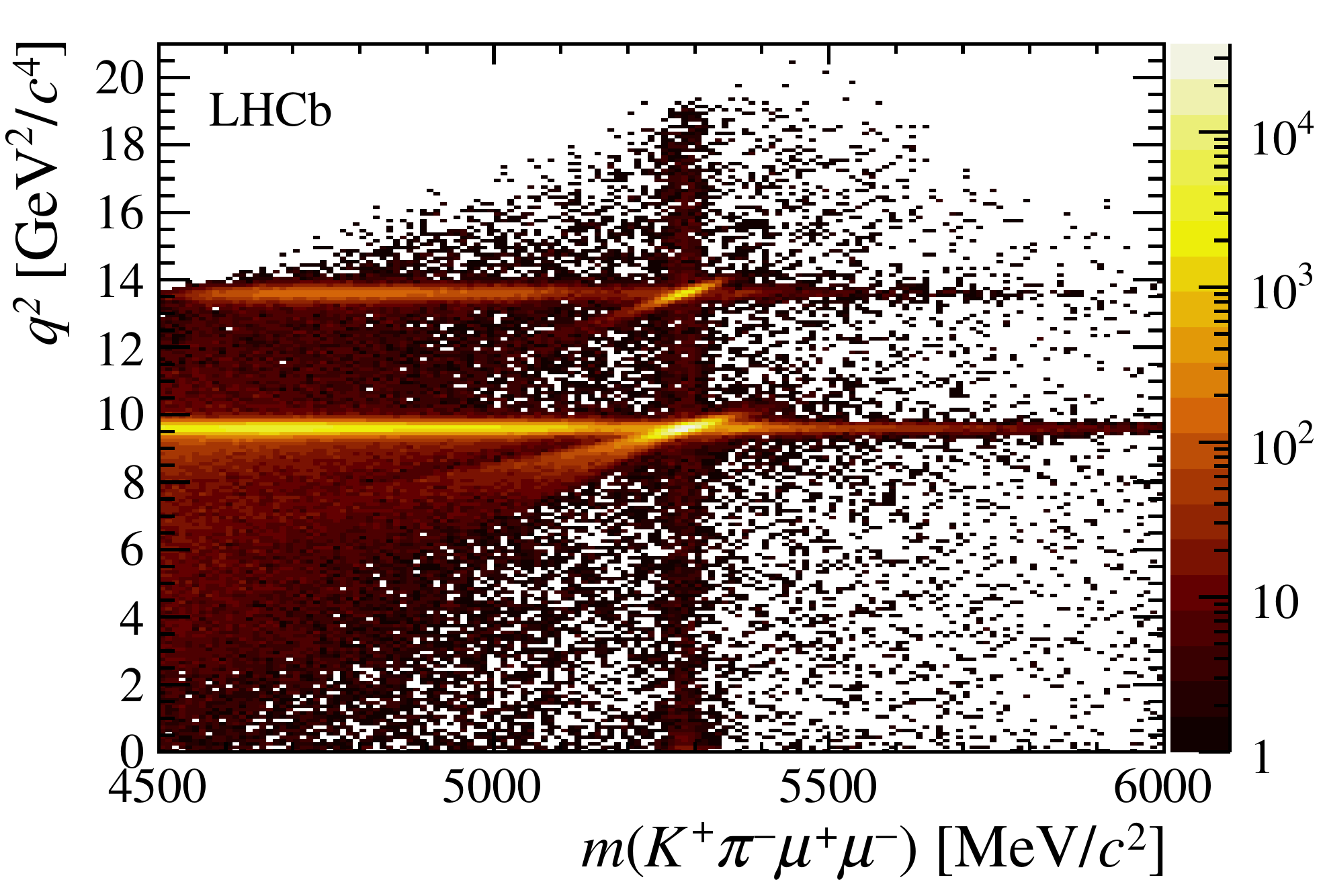} 
\includegraphics[width=0.49\textwidth]{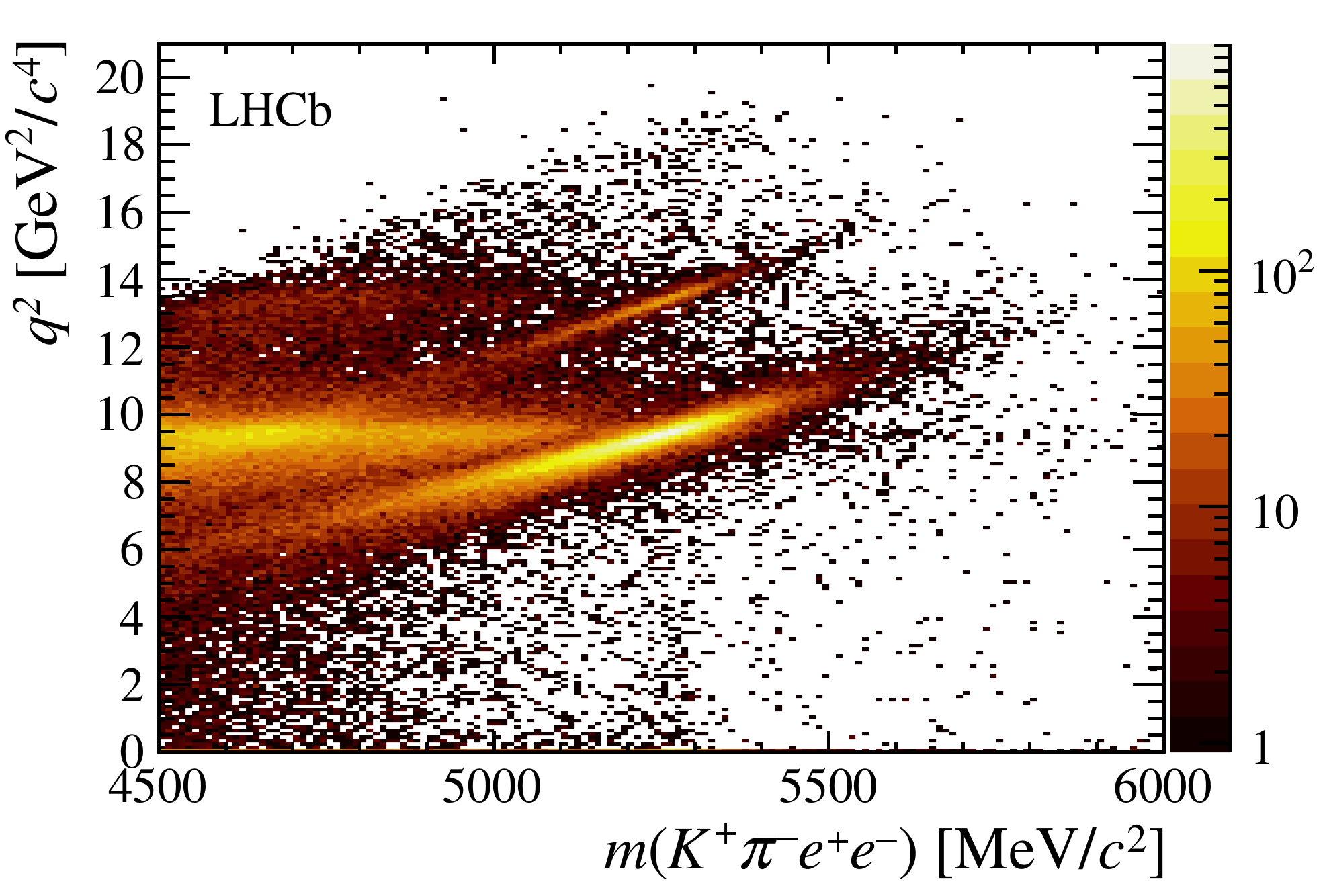}
\caption{Number of candidates for \BdToKstll final states with (left) muons and (right) electrons as a function of the dilepton invariant mass squared, \qsq, and the four-body invariant mass of the \Bz.}
\label{fig:q2vsB}
\end{figure}

A \Bz candidate is formed from a pair of well-reconstructed oppositely charged particles identified as either muons or electrons, combined with two well-reconstructed oppositely charged particles, one identified as a kaon and the other as  a pion.
The \KPi invariant mass is required to be within 100\mevcc of the known \Kstarz mass. 
The kaon and pion must have \pt exceeding 250\mevc, while for the muons (electrons) $\pt > 800 \, (500)\mevc$ is required.
Only dilepton pairs with a good-quality vertex are used to form signal candidates. 
The \Kstarz meson and \lplm pair are required to originate from a common vertex in order to form a \Bz candidate. 
When more than one PV is reconstructed, the one with the smallest \chisqip is selected, where \chisqip is the difference in \chisq of a given PV reconstructed with and without the considered \Bz candidate.
With respect to this selected PV, the impact parameter of the \Bz candidate is required to be small, its decay vertex significantly displaced, and the momentum direction of the \Bz is required to be consistent with its direction of flight.
This direction is given by the vector between the PV and decay vertex. 
The distribution of \qsq as a function of the four-body invariant mass for the \Bz candidates is shown in figure~\ref{fig:q2vsB} for both muon and electron final states. 
The requirements on the neural-network classifier and \mcorr (see section~\ref{sec:selection}) are not applied.
In each plot, the contributions due to the charmonium resonances are clearly visible at the \jpsi and \psitwos masses.
For electrons, these distributions visibly extend above the nominal mass values due to the calorimeter resolution affecting the bremsstrahlung recovery procedure (see section~\ref{sec:eleRecoEff}).
The empty region in the top left corresponds to the kinematic limit of the \BdToKstll decay, while the empty region in the top right corresponds to the requirement that rejects the \BuToKll background (see section~\ref{sec:backgrounds}).
 
The \Bz mass resolution and the contributions of signal and backgrounds depend on the way in which the event was triggered.
The data sample of decay modes involving an \epem pair is therefore divided into three mutually exclusive categories, which in order of precedence are: candidates for which one of the electrons from the \Bz decay satisfies the hardware electron trigger (\loe), candidates for which one of the hadrons from the \Kstarz decay meets the hardware hadron trigger (\loh) requirements, and candidates triggered by activity in the event not associated with any of the signal decay particles (\loi). 
For \BdToKstmm candidates, at least one of the two leptons must satisfy the requirements of the hardware muon trigger.

For the \BdToKstJPsmm decay mode, a dimuon mass interval within 100\mevcc of the known \jpsi mass is selected to identify candidates. 
It is not possible to apply a tight \qsq requirement to identify the \BdToKstJPsee mode as, despite the bremsstrahlung recovery, the \epem invariant mass distribution has a long radiative tail towards low values.
This tail can be seen in figure~\ref{fig:q2vsB}.
The \qsq interval used to select \BdToKstJPsee candidates is between 6.0 and 11.0\gevgevcccc, with the lower limit corresponding to the upper boundary of the \cqsq bin.

The separation of the signal from the combinatorial background is based on neural-network classifiers~\cite{Feindt:2006pm}.
The same classifier is used for the resonant and nonresonant modes, but muon and electron channels are treated separately.
The classifiers are trained using simulated \BdToKstll decays, which have been corrected for known differences between data and simulation (see section~\ref{sec:weights}), to represent the signal.
Data candidates with \KPill invariant masses larger than 5400\mevcc and 5600\mevcc are used to represent background samples for the muon and electron channel, respectively. 
To best exploit the size of the available data sample for the training procedure, a $k$-folding technique~\cite{Blum:1999:BHB:307400.307439} is adopted with $k=10$.
The variables used as input to the classifiers are: the transverse momentum, the quality of the vertex fit, the \chisqip, the \chisqfd (the \chisq on the measured distance between the PV and the decay vertex), and the angle between the direction of flight and the momentum of the \Bz candidate, the \KPi and the dilepton pairs;
the minimum and maximum of the kaon and pion \pt, and of their \chisqip;
the minimum and maximum of the lepton \pt values, and of their \chisqip;
and finally, the most discriminating variable, the quality of the kinematic fit to the decay chain (this fit is performed with a constraint on the vertex that requires the \Bz candidate to originate from the PV).
In each fold, only variables that significantly improve the discriminating power of the classifier are kept.

\begin{figure}[t!]
\centering
\includegraphics[width=.8\textwidth]{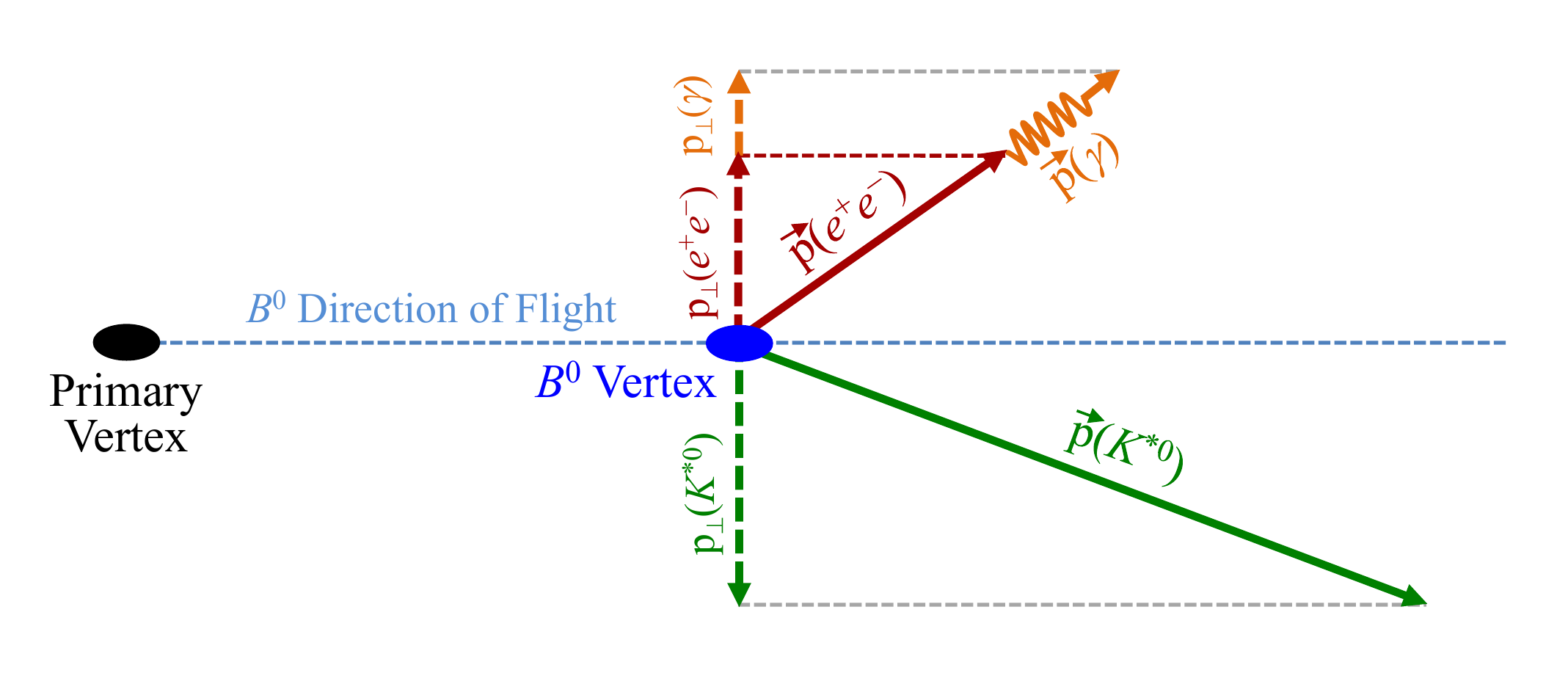} 
\caption{Sketch of the topology of a \BdToKstee decay. The transverse momentum lost via bremsstrahlung is evaluated as the difference between the \pt of the \Kstarz meson and that of the dielectron system, where both are calculated with respect to the \Bz meson direction of flight. Bremsstrahlung photons that are not recovered by the reconstruction are assumed to follow the dielectron momentum direction.}
\label{fig:HOP}
%\end{figure}
\vspace{0.5cm}
%\begin{figure}[t!]
\centering
\includegraphics[width=0.49\textwidth]{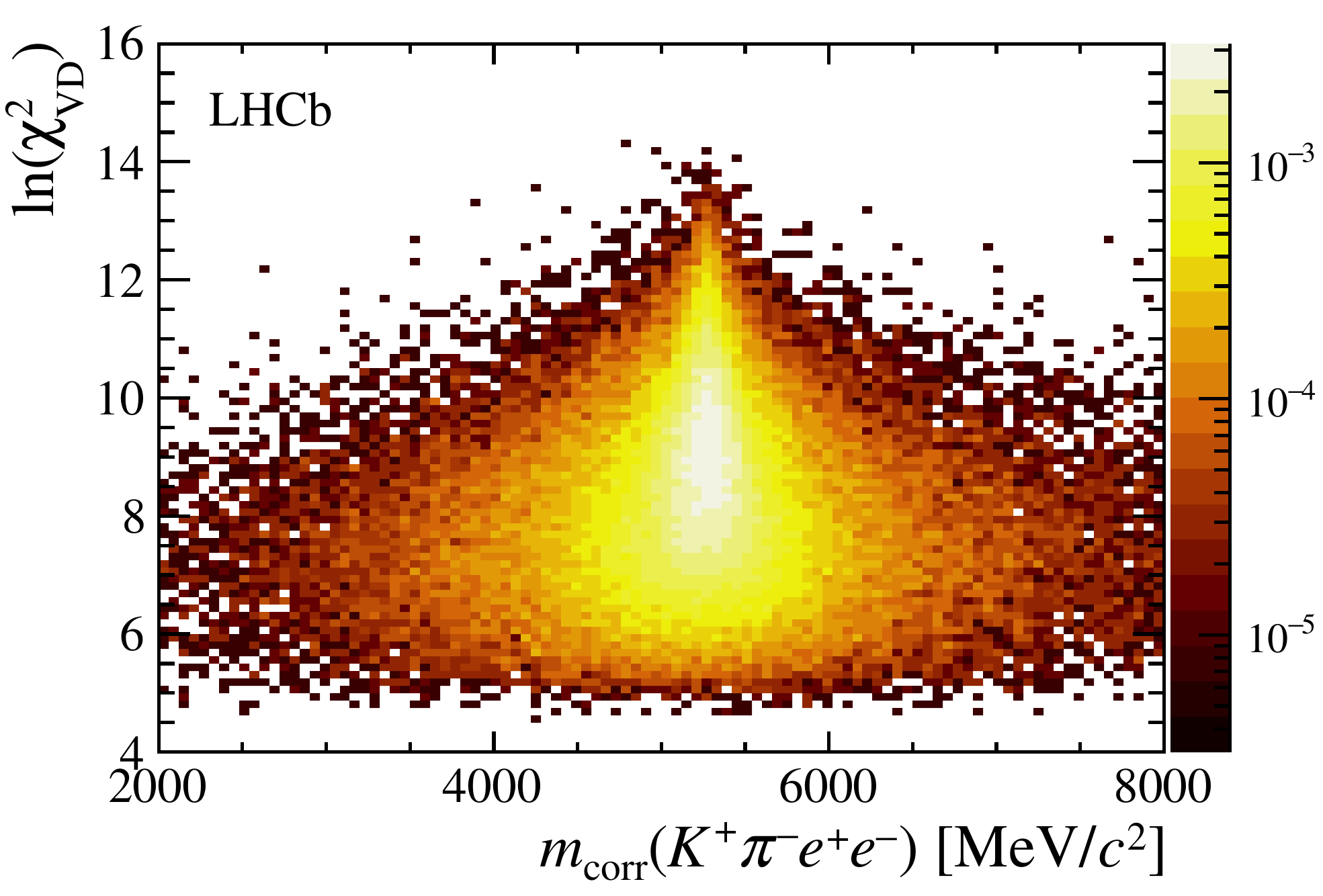}
\includegraphics[width=0.49\textwidth]{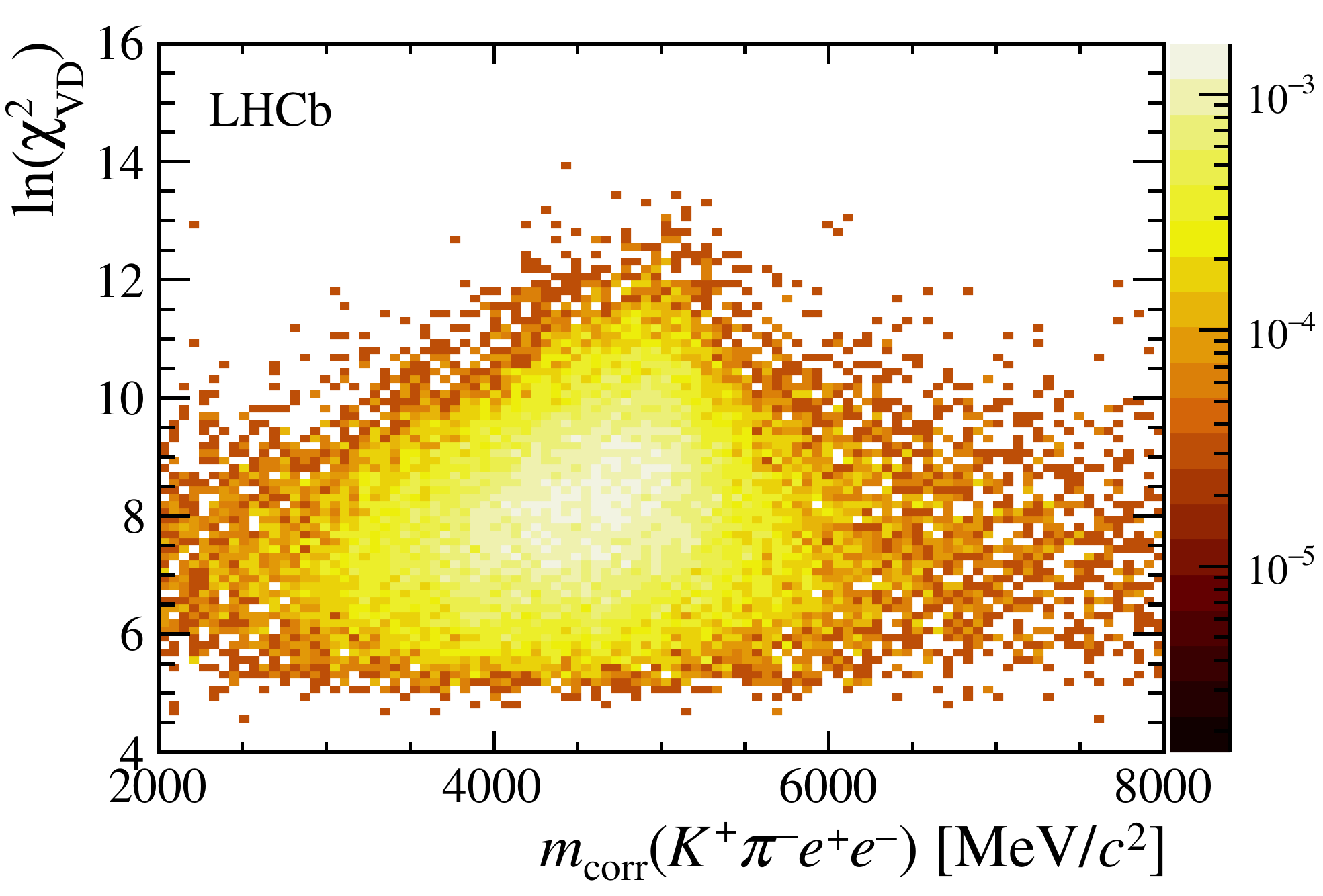}
\caption{Distribution of \chisqfd as a function of \mcorr for (left) \BdToKstee and (right) \BToXee simulated candidates. The distributions are normalised to the same number of candidates.}
\label{fig:HOP2D}
\end{figure}

For the muon modes, a requirement on the four-body invariant mass of the \Bz candidate to be larger than 5150\mevcc excludes backgrounds due to partially reconstructed decays, \BToXmm, where one or more of the products of the \B decay, denoted as $X$, are not reconstructed.
A kinematic fit that constrains the dielectron mass to the known \jpsi mass allows the corresponding background to be separated from the \BdToKstJPsee signal by requiring the resulting four-body invariant mass to be at least 5150\mevcc.
For the nonresonant electron mode, the partially reconstructed backgrounds can be reduced by exploiting the kinematics of the decay. 
The ratio of the \Kstarz and the dielectron momentum components transverse to the \Bz direction of flight is expected to be unity, unless the electrons have lost some energy due to bremsstrahlung that was not recovered (see figure~\ref{fig:HOP}).
In the approximation that bremsstrahlung photons do not modify the dielectron direction significantly, which is particularly valid for low dilepton masses, this ratio can be used to correct the momentum of the dielectron pair.
The invariant mass of the signal candidate calculated using the corrected dielectron momentum, \mcorr, has a poor resolution that depends on \chisqfd.
Nevertheless, since the missing momentum of background candidates does not originate from the dielectron pair, \mcorr still acts as a useful discriminating variable.
Signal and partially reconstructed backgrounds populate different regions of the two-dimensional plane defined by \mcorr and \chisqfd (see figure~\ref{fig:HOP2D}).
The requirements in this plane and on the classifier response are optimised simultaneously, but separately for each \qsq region.
The optimisation maximises a figure of merit defined as $N_S / \sqrt{N_S+N_B}$, where the expected signal yield, $N_S$, is evaluated by scaling the observed number of \BdToKstJPsll candidates by the ratio of the branching fractions of the nonresonant and resonant modes, and the expected background yield, $N_B$, is obtained by fitting the mass sidebands in data.

After the full selection, 1 to 2\% of the events contain multiple candidates.
This fraction is consistent between the resonant and nonresonant modes, and between final states with electrons and muons.
About half of the multiple candidates are due to cases where the kaon is misidentified as the pion and vice versa.
In all cases only one candidate, chosen randomly, is retained.

\section{Exclusive backgrounds}
\label{sec:backgrounds}

Specific requirements are applied to reject backgrounds from \bquark-hadron decays, while ensuring a negligible loss of signal, as verified using simulation.
In the \lqsq region, the size of the contamination from \BdToKstVll decays, where $V$ is a $\rho$, $\omega$ or $\phi$ meson, is evaluated in Refs.~\cite{Korchin:2010uc, Jager:2012uw}.
The contamination due to direct decays or interference with the signal channel is found to be smaller than 2\% and similar for muons and electrons.
As a consequence, the residual effect in the double ratio is expected to be very small and can therefore be safely neglected.

Misreconstructed \BdToKstJPsmm and \BdToKstPsimm decays can contaminate the signal region if the identities of one of the hadrons and one of the muons are swapped. To avoid this, the invariant mass of the hadron candidate (under the muon mass hypothesis) and the oppositely charged muon is required to be outside of a 60\mevcc interval around the known \jpsi or the \psitwos masses.

A large, nonpeaking background comes from the \mbox{\decay{\Bd}{\Dm \lp \nu}} decay, with \mbox{\decay{\Dm}{\Kstarz \ln \overline{\nu}}}, which has a 
branching fraction four orders of magnitude larger than that of the signal.
In the rare case where both neutrinos have low energies, the signal selection will be less effective at rejecting this background.
This decay can be separated from the signal by exploiting the angular distribution of the dilepton pair.
For \mbox{\decay{\Bd}{\Dm \lp \nu}} decays, the angle $\theta_\ell$ between the direction of the \lp in the dilepton rest frame and the direction of the dilepton in the \Bz rest frame tends to be small.
This background is suppressed by requiring \mbox{$|\ctl|<0.8$}. 

When combined with a low-momentum \pim meson from the rest of the event, \BuToKll  decays can pass the selection and populate the upper mass sideband region that is used to represent the combinatorial background for the training of the neural-network classifiers.
Such decays are vetoed by requiring the invariant mass of the $\Kp\ellell$ combination to be less than \mbox{5100\mevcc}.
Candidates where the \pim from the \Kstarz is misidentified as a kaon and paired with a \pip are similarly rejected.
To suppress background from \BsToPhill decays, with \mbox{\decay{\phi}{\Kp\Km}} where one of the kaons is misidentified as a pion, the invariant mass of the two hadrons computed under the $\Kp\Km$ mass hypothesis is required to be larger than 1040\mevcc.

%%%%%%%%%%%%%%%%%%%%%%%%%%%%%%%%%%%%
% !TEX root = main.tex
%%%%%%%%%%%%%%%%%%%%%%%%%%%%%%%%%%%%

\section{Fits to the \boldmath{\KPill} invariant mass distributions}
\label{sec:yields}

The signal yields are determined using unbinned extended maximum likelihood fits to the four-body invariant mass, \mKpill, of the selected candidates in each \qsq interval and for each lepton type. 
The reconstructed invariant mass is calculated using a kinematic fit with a constraint on the vertex that requires the \Bz candidate to originate from the PV. 
In order to improve the quality and stability of the results, the fits are performed simultaneously on the nonresonant and resonant modes, and some parameters are shared.

For the muon channel, the fit is performed in an invariant mass window of \mbox{5150--5850\mevcc}.
The low edge is chosen to reject the partially reconstructed background that populates the low mass region. 
The probability density function (PDF) for the signal is defined by a Hypatia function~\cite{Hypatia}, where the parameters are fixed from simulation.
However, in order to account for possible residual discrepancies with data, the mean and width are allowed to vary freely in the fit, independently for the resonant and nonresonant modes and in each \qsq region. 
The combinatorial background is parameterised using an exponential function, which has a different slope in the resonant and nonresonant modes, and in each \qsq region, that is free to vary in the fit.
For the resonant mode, two additional sources of background are included: \LbTopKJPsmm decays, where the \antiproton candidate is misidentified as a \pim meson, and \BsToKstJPsmm decays.
The former are described using a kernel estimation technique~\cite{Cranmer:2000du} applied to simulated events for which the \KPi invariant mass distribution has been matched to data from Ref.~\cite{LHCb-PAPER-2015-029}.
The latter are modelled using the same PDF as for the signal, but with the mean value shifted by the known difference between the \Bz and the \Bs masses. 
The equivalent backgrounds to the nonresonant mode are found to be negligible.

\begin{figure}[t!]
\centering
\includegraphics[width=0.49\textwidth]{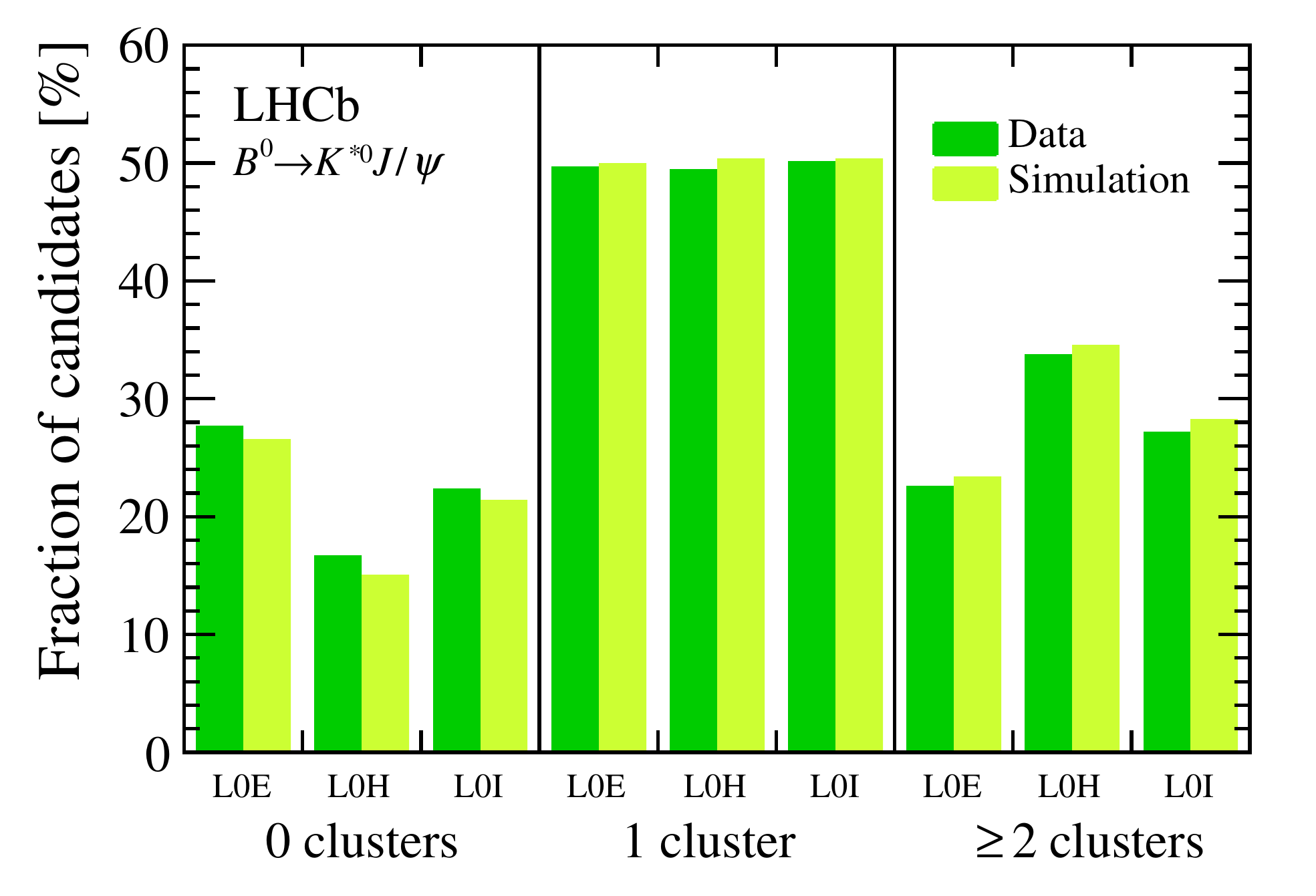}
\includegraphics[width=0.49\textwidth]{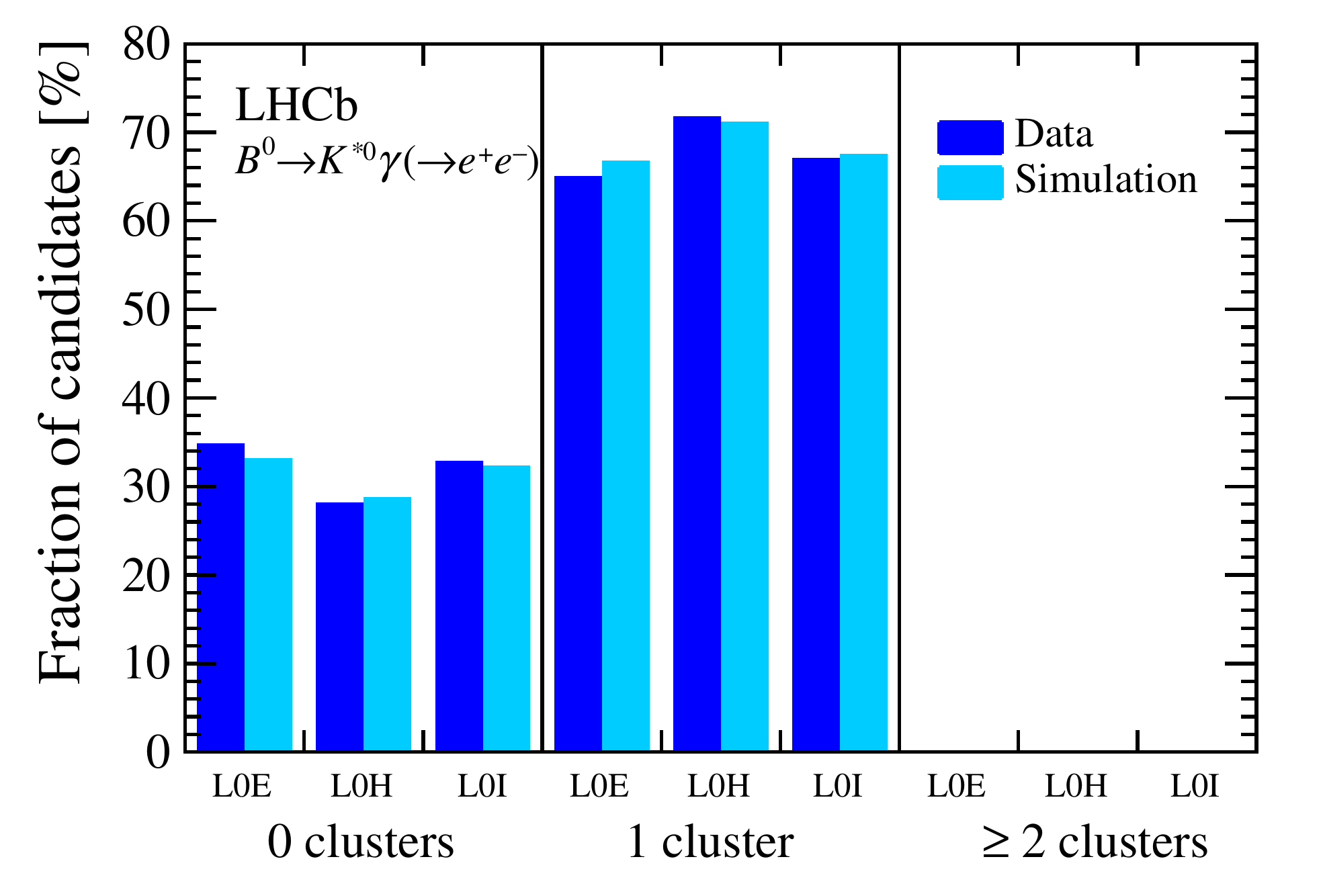}
\caption{Fraction of (left) \BdToKstJPsee and (right) \BdToKstGee candidates (in percent) with zero, one, and two or more recovered clusters per trigger category. The numbers are from (darker colour) data and (lighter colour) simulation. Due to the very low opening angle of the two electrons in \BdToKstGee decays, 
the bremsstrahlung photon energy deposits overlap and only one bremsstrahlung cluster at most is resolved.}
\label{fig:brem}
\end{figure}

For the electron channel, due to the  limited resolution on the \KPiee invariant mass,  a wider  window of 4500--6200\mevcc is used. 
The resolution on the reconstructed invariant mass of the \Bz and the background composition depends on the kinematics of the decay, as well as on the trigger category. 
For this reason, simultaneous fits to the four-body invariant mass of the \BdToKstJPsee and \BdToKstee channels are performed separately in the three trigger categories. 
Following the strategy of Ref.~\cite{LHCb-PAPER-2014-024}, the \KPiee signal PDF is observed to depend on the number of calorimeter clusters that are added to the dielectron candidate in order to correct for the effects of bremsstrahlung.
Three bremsstrahlung categories are considered, depending on whether zero, one or more clusters are recovered.
The PDF is described by the sum of a Crystal Ball function~\cite{Skwarnicki:1986xj} (CB) and a wide Gaussian function.
The CB function accounts for FSR and bremsstrahlung that is not fully recovered, and corresponds to over 90\% of the total signal PDF.
Cases where bremsstrahlung clusters were incorrectly associated are accounted for by the Gaussian function.
The shape parameters and the fraction of candidates in each bremsstrahlung category are taken from simulation, the latter having been checked on data control channels (see figure~\ref{fig:brem}). 
In order to account for possible data-simulation discrepancies, the mean (width) of the PDF for each trigger category is allowed to shift (scale). 
These shift and scale factors are common between the nonresonant and resonant PDFs.
An additional scale factor is also applied to the parameter describing the tail of the CB functions. 
The combinatorial background is described by an exponential function with different slope parameters for the resonant and nonresonant modes, and in each trigger category and \qsq region, that are free to vary in the fit.
The shape of the partially reconstructed hadronic background, \BToXee (where the decay product $Y$ is not reconstructed), is obtained from simulation using a sample that includes decays of higher kaon resonances, $X$, such as \Kone and \Ktwo.
The mass distribution is modelled using a kernel estimation technique separately in each trigger category and \qsq region.
The fraction of this background is free to vary in both \qsq intervals. 
Due to the requirement on the four-body invariant mass with a \jpsi mass constraint (see section~\ref{sec:selection}), there is no partially reconstructed background left to contaminate \BdToKstJPsee candidates.
Due to the long radiative tail of the dielectron invariant mass, \BdToKstJPsee decays can contaminate the \cqsq region and an additional background component is considered (see figure~\ref{fig:q2vsB}), however this contribution does not peak at the nominal \Bz mass.
The distribution is modelled using simulated events, while the normalisation is constrained using a mixture of data and simulation.
The contributions to the resonant modes from \LbTopKJPsee and  \BsToKstJPsee decays are treated following the same procedure as for the muon channel.
The normalisations are fixed to the yields returned by the muon fit after correcting for efficiency differences between the two final states. 

\begin{figure}[t!]
\centering
\includegraphics[width=0.49\textwidth]{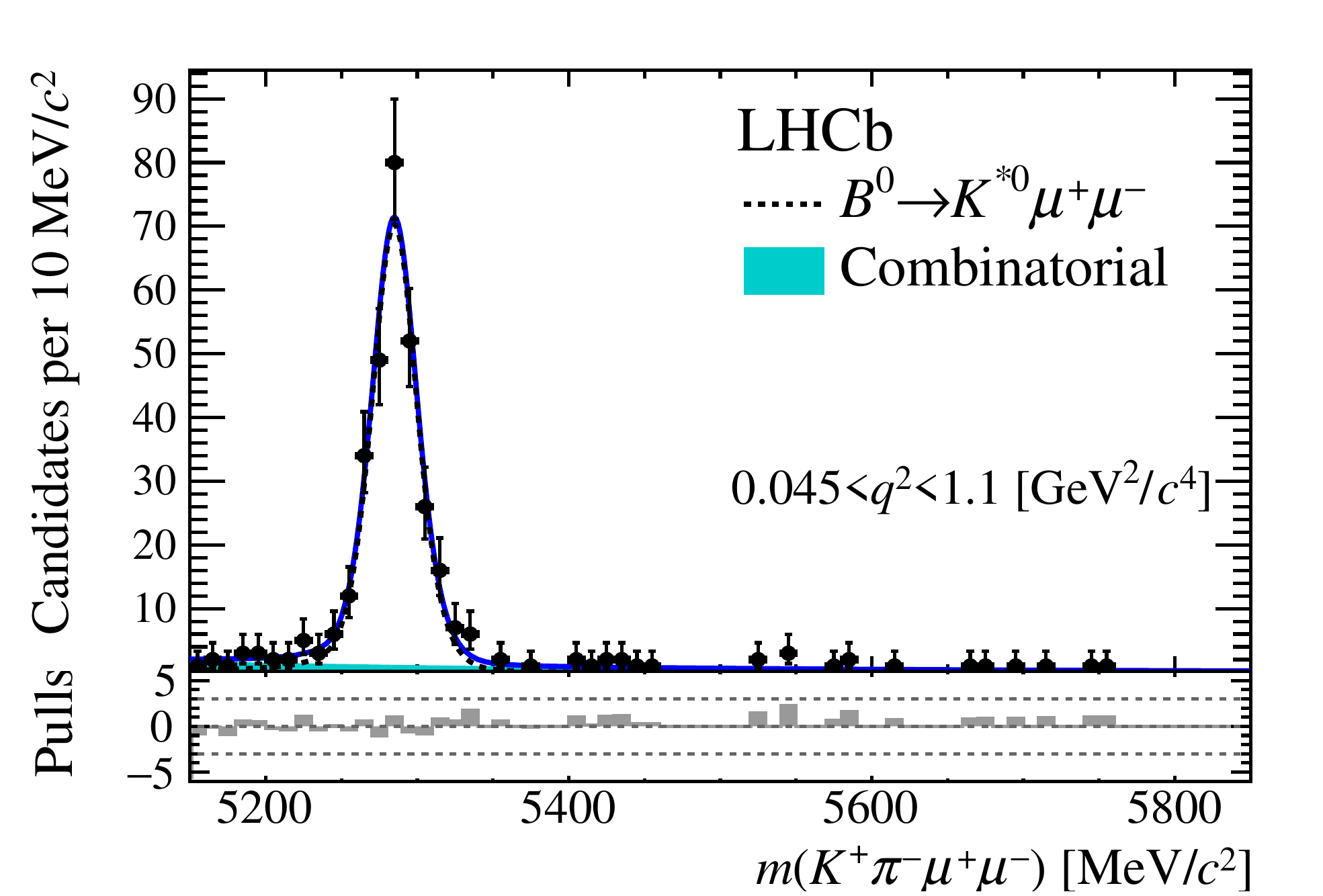}
\includegraphics[width=0.49\textwidth]{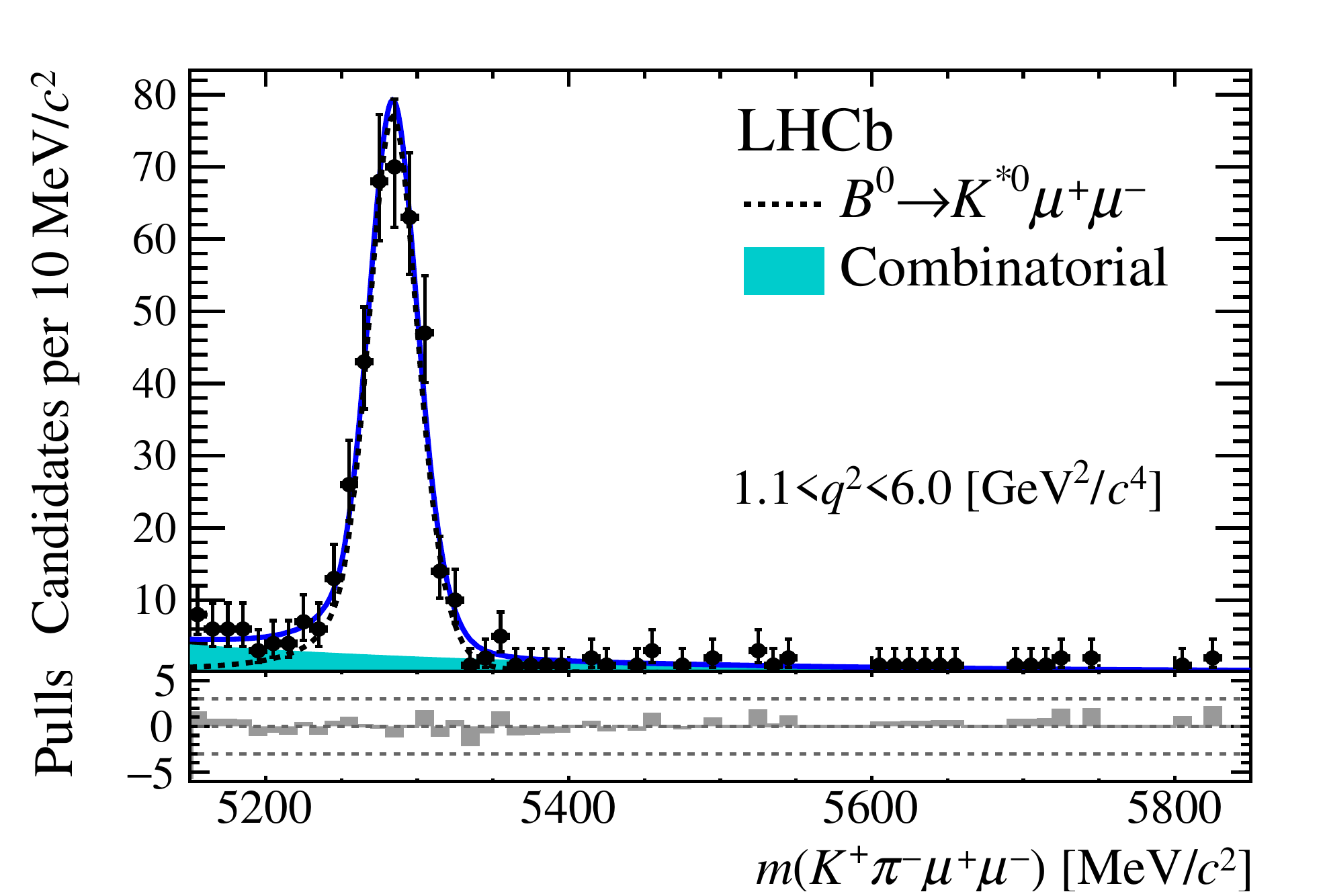}
\includegraphics[width=0.49\textwidth]{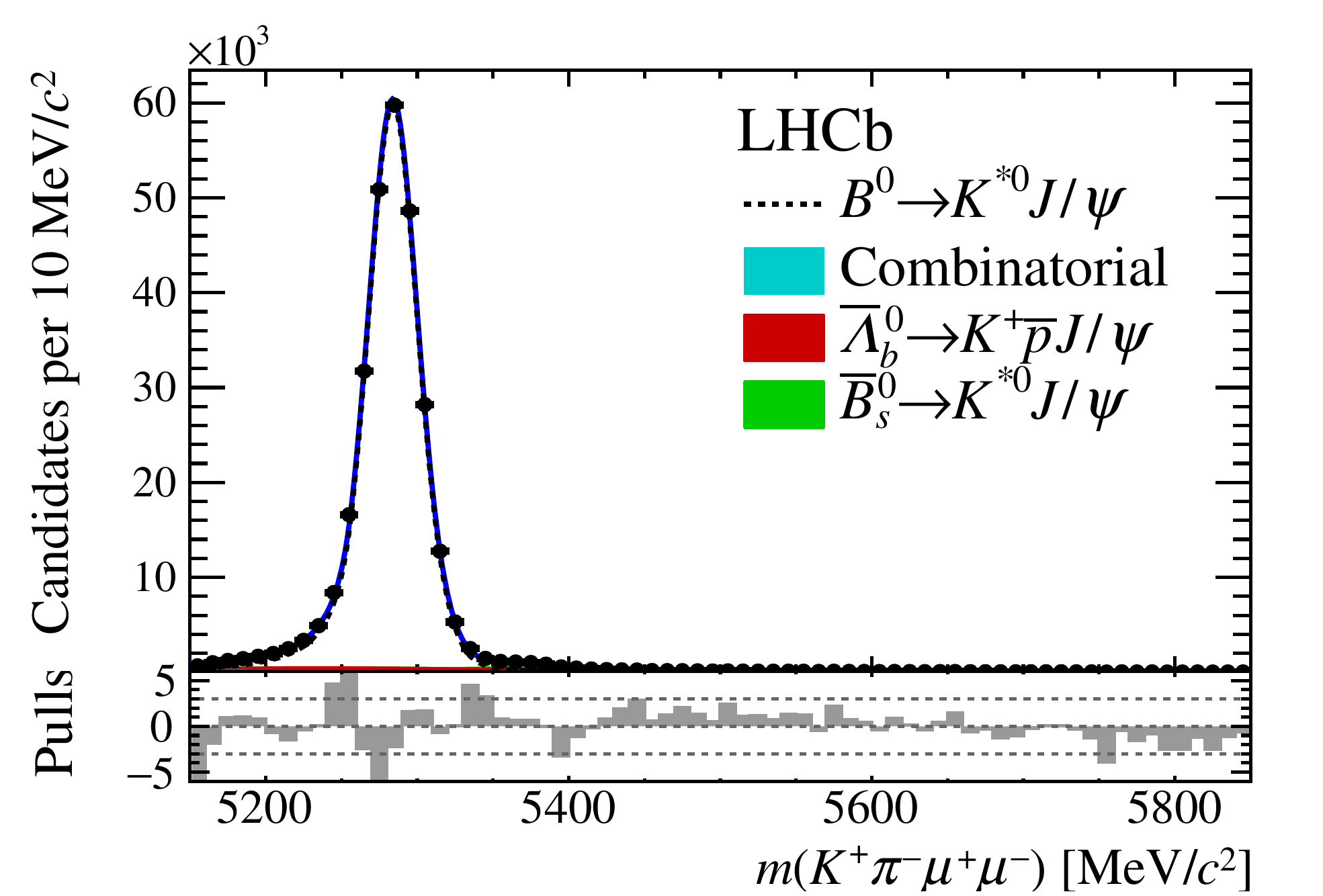}
\caption{Fit to the \mKpimm invariant mass of (top) \BdToKstmm in the low- and \cqsq bins and (bottom) \BdToKstJPsmm candidates. The dashed line is the signal PDF, the shaded shapes are the background PDFs and the solid line is the total PDF. The fit residuals normalised to the data uncertainty are shown at the bottom of each distribution.}
\label{fig:fitMM}
\end{figure}

\begin{figure}[t!]
\centering
\includegraphics[width=0.49\textwidth]{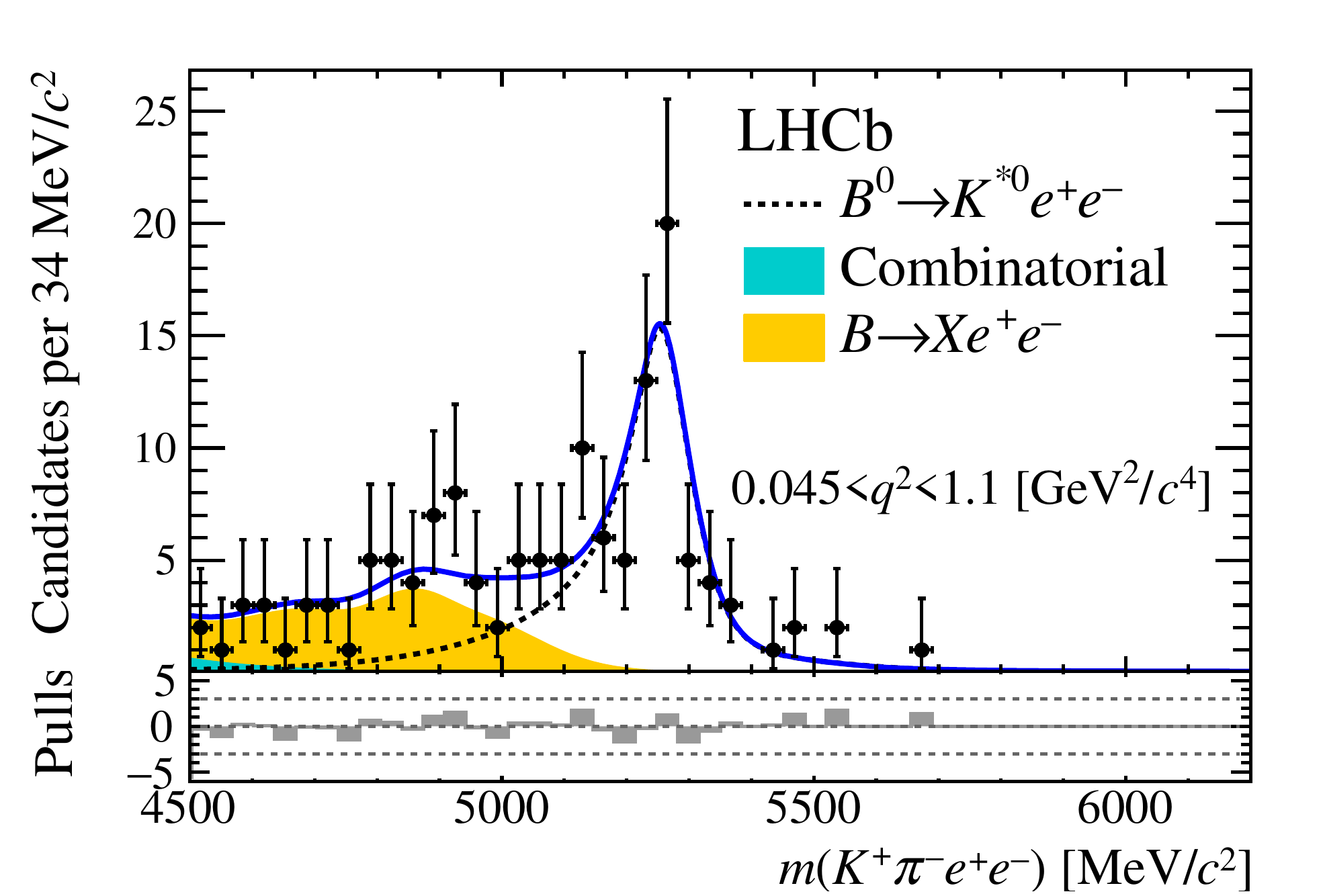}
\includegraphics[width=0.49\textwidth]{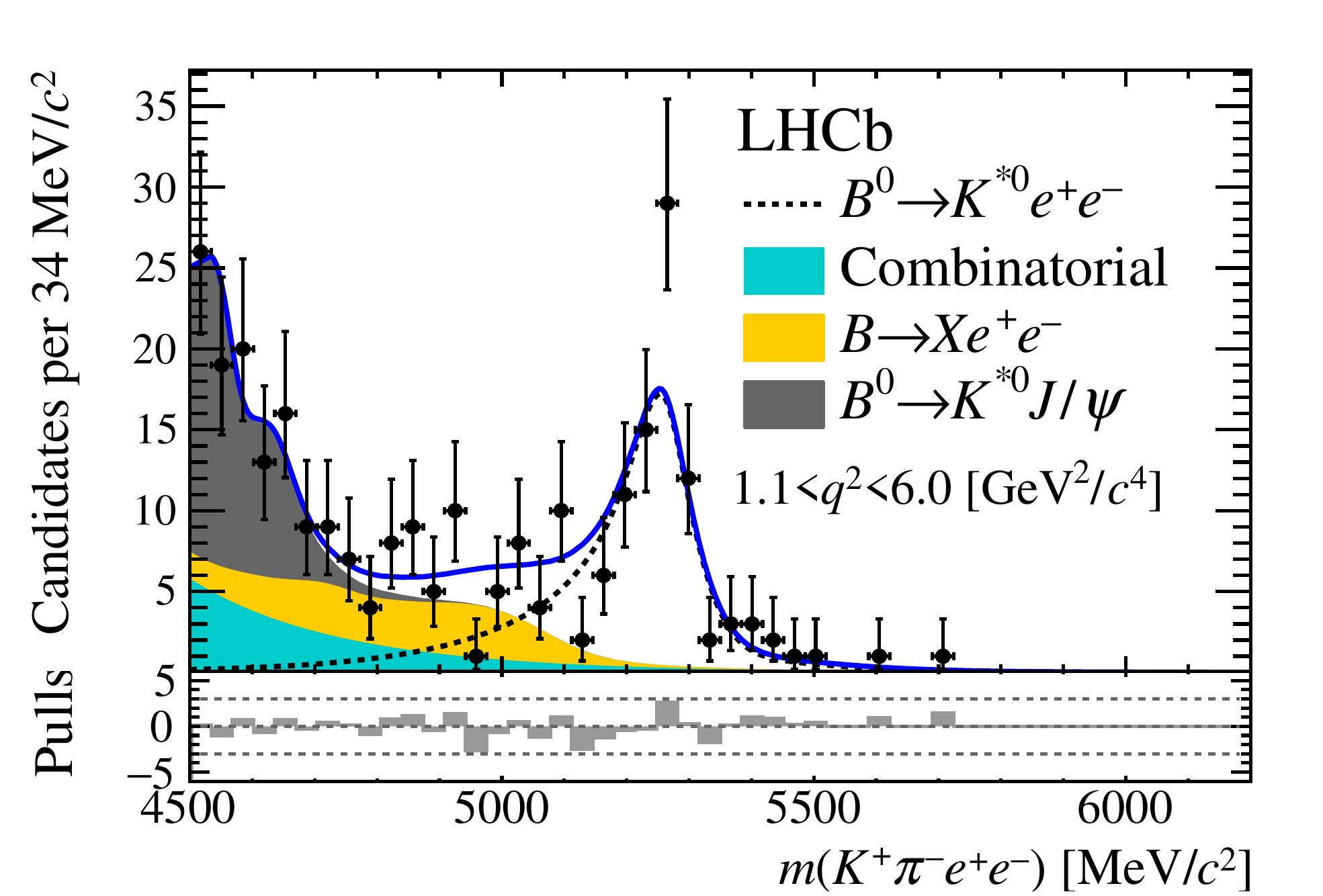}
\includegraphics[width=0.49\textwidth]{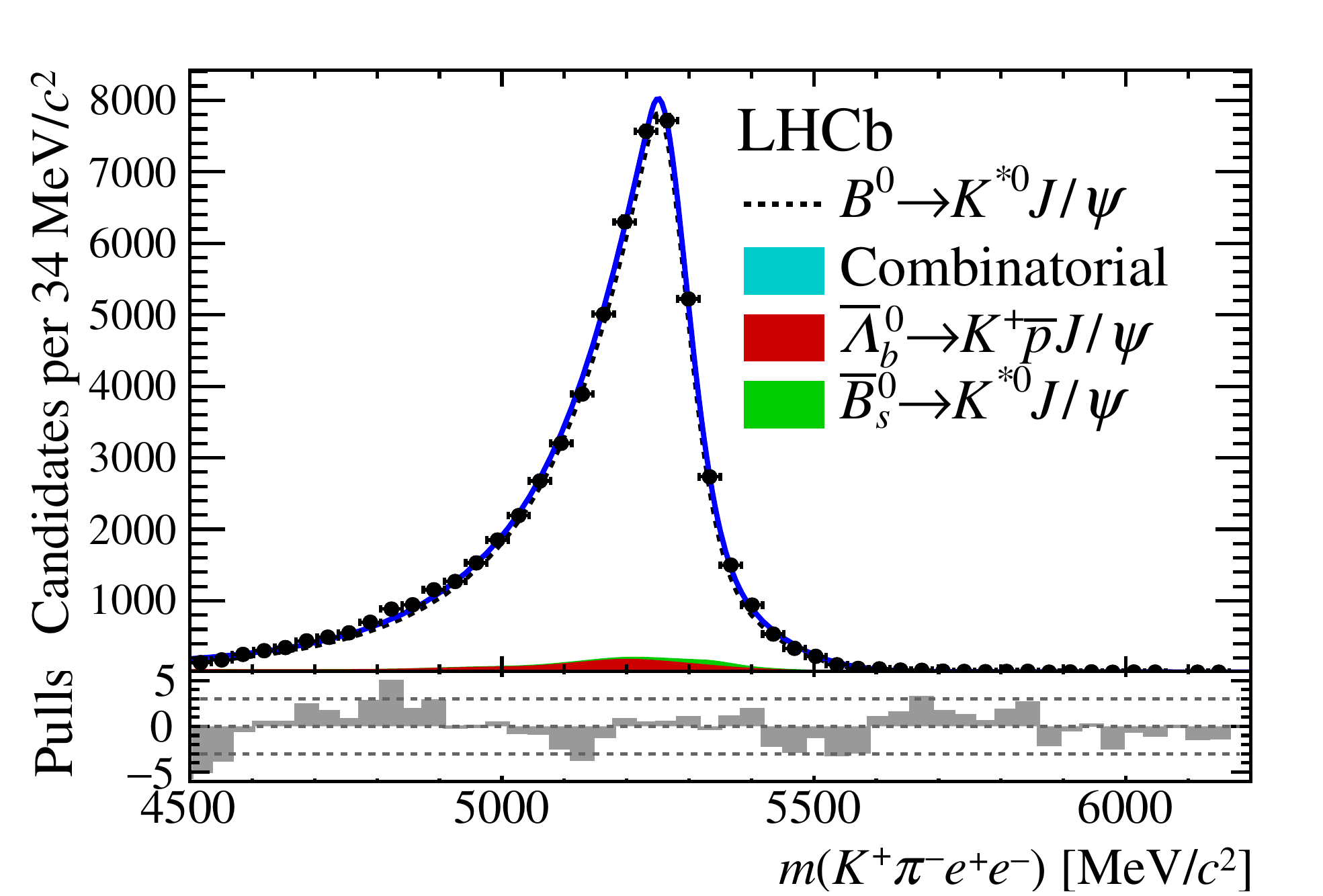}
\caption{Fit to the \mKpiee invariant mass of (top) \BdToKstee in the low- and \cqsq bins and (bottom) \BdToKstJPsee candidates. The dashed line is the signal PDF, the shaded shapes are the background PDFs and the solid line is the total PDF. The fit residuals normalised to the data uncertainty are shown at the bottom of each distribution.}
\label{fig:fitEE}
\end{figure}

The results of the fits to the muon channels are shown in figure~\ref{fig:fitMM}, while figure~\ref{fig:fitEE} displays the fit results for the electron channels, where the three trigger categories have been combined.
The distribution of the normalised fit residuals of the \BdToKstJPsmm mode shows an imperfect description of the combinatorial background at high mass values, although the effect on the signal yield is negligible.
The resulting yields are listed in table~\ref{tab:yields}. 

\begin{table}[t!]
\centering
\caption{Yields obtained from the mass fits to the muon and electron (in the three trigger categories) channels. The uncertainties are statistical only.} 
\label{tab:yields}
\renewcommand\arraystretch{1.4}
\begin{tabular}{c|c|c|c}
			& \multicolumn{2}{c|}{\BdToKstll}						& \multirow{2}{*}{\BdToKstJPsll} \\ 
\cline{2-3}
			& \lqsq						& \cqsq	& \\ \hline
\mpmm		&  $285~^{+~18}_{-~18}$			&  $353~^{+~21}_{-~21}$	& $274416~^{+~602}_{-~654}$ \\ \hline
\epem (\loe)	&  $\phantom{0}55~^{+~\phantom{0}9}_{-~\phantom{0}8}$	& $\phantom{0}67~^{+~10}_{-~10}$		& $\phantom{0}43468~^{+~222}_{-~221}$ \\
\epem (\loh)	&  $\phantom{0}13~^{+~\phantom{0}5}_{-~\phantom{0}5}$	& $\phantom{0}19~^{+~\phantom{0}6}_{-~\phantom{0}5}$		& $\phantom{00}3388~^{+~\phantom{0}62}_{-~\phantom{0}61}$ \\
\epem (\loi)	&  $\phantom{0}21~^{+~\phantom{0}5}_{-~\phantom{0}4}$	& $\phantom{0}25~^{+~\phantom{0}7}_{-~\phantom{0}6}$		& $\phantom{0}11505~^{+~115}_{-~114}$ \\
\end{tabular}
\end{table}

%%%%%%%%%%%%%%%%%%%%%%%%%%%%%%%%%%%%
% !TEX root = main.tex
%%%%%%%%%%%%%%%%%%%%%%%%%%%%%%%%%%%%

\section{Efficiencies}
\label{sec:eff}

The efficiency for selecting each decay mode is defined as the product of the efficiencies of the geometrical acceptance of the detector, the complete reconstruction of all tracks, the trigger requirements and the full set of kinematic, PID and background rejection requirements. 
All efficiencies are determined using simulation that is tuned to data, as described in section~\ref{sec:weights}, and account for bin migration in \qsq due to resolution, FSR and bremsstrahlung in the detector.
The net bin migration amounts to about 1\% and 5\% in the low- and central-\qsq regions, respectively.

The efficiency ratios between the nonresonant and the resonant modes, $\varepsilon_{\ellell} / \varepsilon_{\JPsll}$, which directly enter in the \RKst measurement, are reported in table~\ref{tab:efficiencyRatio}.
Besides a dependence on the kinematics, the difference between the ratios in the two \qsq regions is almost entirely due to the different requirement on the neural-network classifier.
The relative fraction of the electron trigger categories is checked using simulation to depend on \qsq as expected: the fraction of \loe decreases when decreasing in \qsq, while \loh increases; on the other hand, the fraction of \loi only mildly depends on \qsq.

\begin{table}[h!]
\centering
\caption{Efficiency ratios between the nonresonant and resonant modes, $\varepsilon_{\ellell} / \varepsilon_{\JPsll}$, for the muon and electron (in the three trigger categories) channels. The uncertainties are statistical only.}
\label{tab:efficiencyRatio}
\renewcommand\arraystretch{1.4}
\begin{tabular}{c|c|c}
			& \multicolumn{2}{c}{$\varepsilon_{\ellell} / \varepsilon_{\JPsll}$} \\
\cline{2-3}
			& \lqsq	& \cqsq \\ \hline
\mpmm		& $0.679 \pm 0.009$				& $0.584 \pm 0.006$ \\ \hline
\epem (\loe)	& $0.539 \pm 0.013$				& $0.522 \pm 0.010$ \\
\epem (\loh)	& $2.252 \pm 0.098$				& $1.627 \pm 0.066$ \\
\epem (\loi)	& $0.789 \pm 0.029$				& $0.595 \pm 0.020$ \\
\end{tabular}
\end{table}

\section{Cross-checks}
\label{sec:crosschecks}

A large number of cross-checks were performed before unblinding the result.
The control of the absolute scale of the efficiencies is tested by measuring the ratio of the branching fractions of the muon and electron resonant channels
\begin{eqnarray*}
\RJPs = \frac{\BR(\BdToKstJPsmm)} {\BR(\BdToKstJPsee)} \, ,
\end{eqnarray*}
which is expected to be equal to unity.
This quantity represents an extremely stringent test, as it does not benefit from the large cancellation of the experimental systematic effects provided by the double ratio.
The \RJPs ratio is measured to be $1.043 \pm 0.006 \pm 0.045$, where the first uncertainty is statistical and the second systematic.
The same sources of systematic uncertainties as in the \RKst measurement are considered (see section~\ref{sec:systematics}).
The result, which is in good agreement with unity, is observed to be compatible with being independent of the decay kinematics, such as \pt and $\eta$ of the \Bz candidate and final-state particles, and the charged-track multiplicity in the event.

The extent of the cancellation of residual systematics in \RKst is verified by measuring a double ratio, \RCC, where \BdToKstPsill decays are used in place of \BdToKstll.
The \RCC ratio, measured with a statistical precision of about 2\%, is found to be compatible with unity within one standard deviation.

The branching fraction of the decay \BdToKstmm is measured and found to be in good agreement with Ref.~\cite{LHCb-PAPER-2016-012}.
Furthermore, the branching fraction of the \BdToKstG decay, where decays with a photon conversion are used, is determined with a statistical precision of about 7\% and is observed to be in agreement with the expectation within two standard deviations.
The \BdToKstGee selection and determination of the signal yield closely follows that of the \BdToKstee decay.

If no correction is made to the simulation, the ratio of the efficiencies changes by less than 5\%.
The relative population of the three bremsstrahlung categories is compared between data and simulation using both \BdToKstJPsee and \BdToKstGee candidates to test possible \qsq dependence of the modelling.
Good agreement is observed, as shown in figure~\ref{fig:brem}.

\begin{figure}[t!]
\centering
\includegraphics[width=0.42\textwidth]{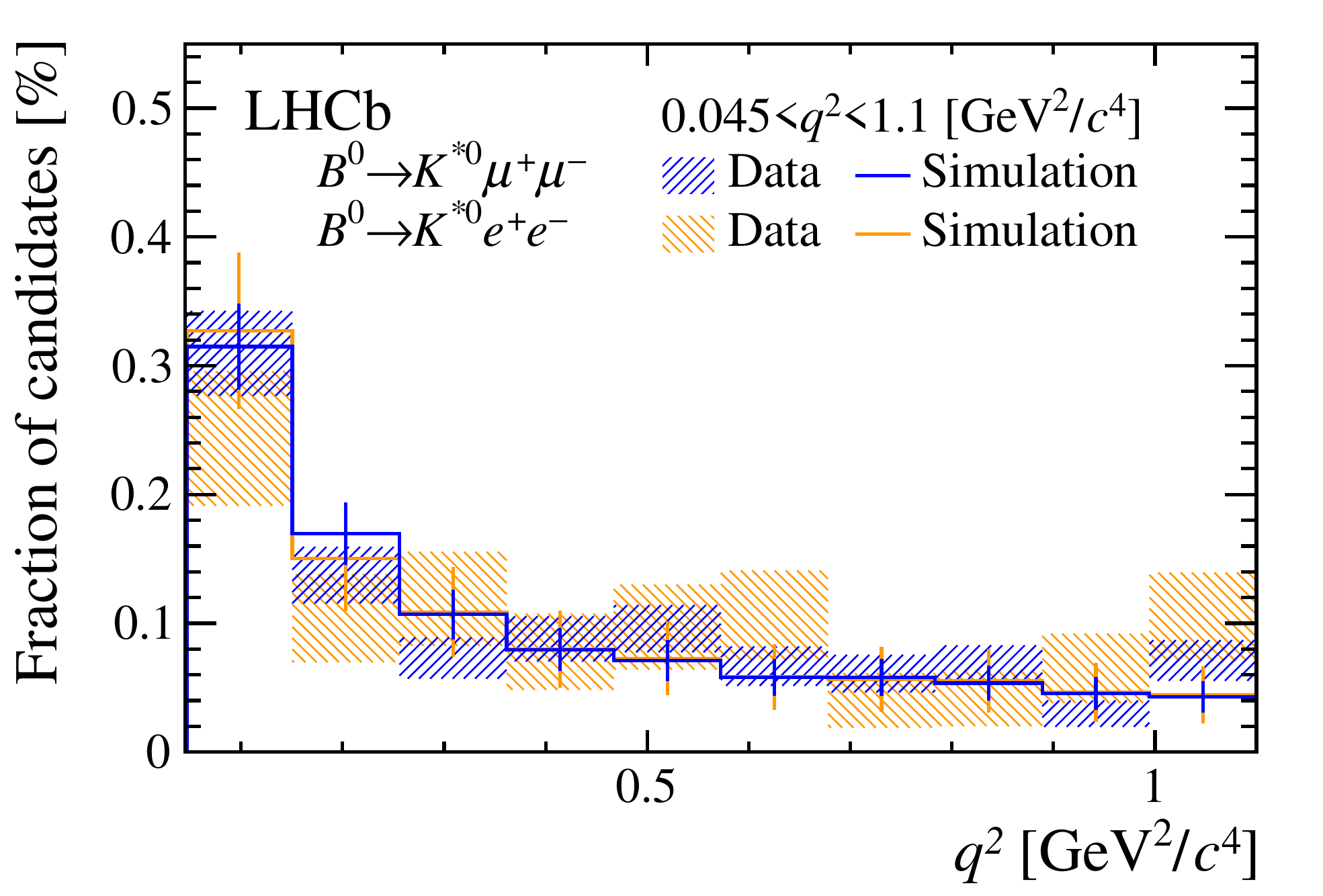}
\includegraphics[width=0.42\textwidth]{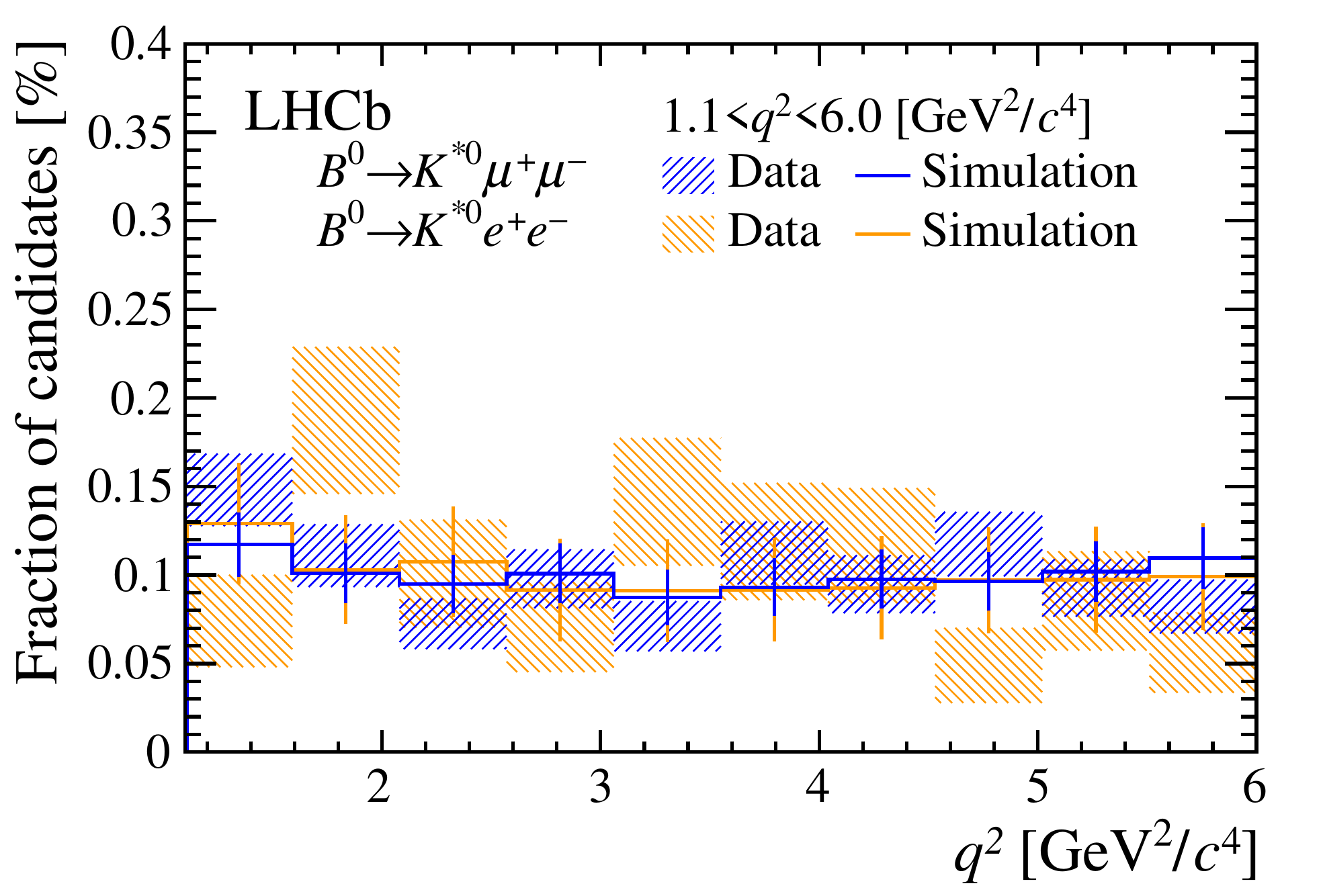}
\includegraphics[width=0.42\textwidth]{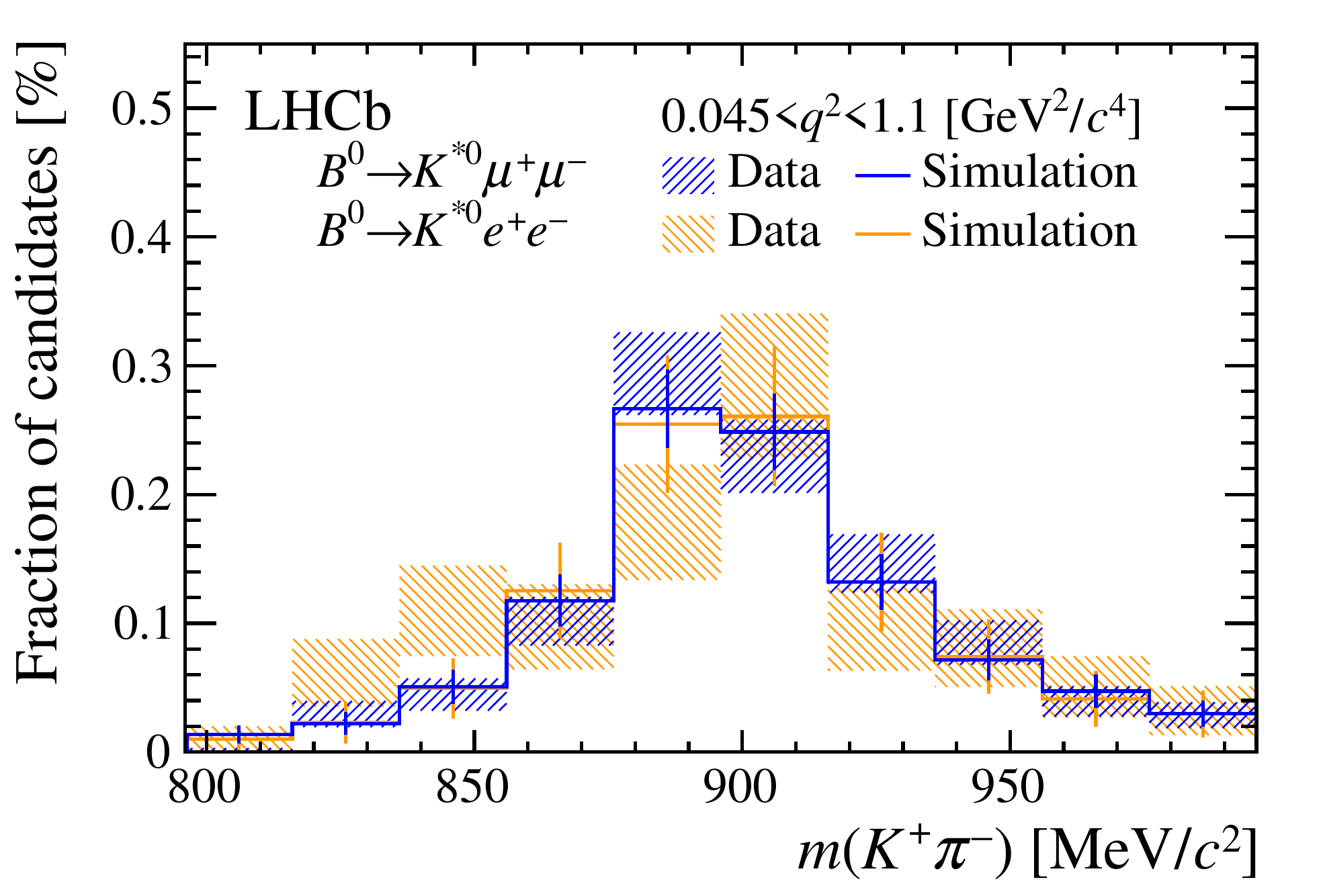}
\includegraphics[width=0.42\textwidth]{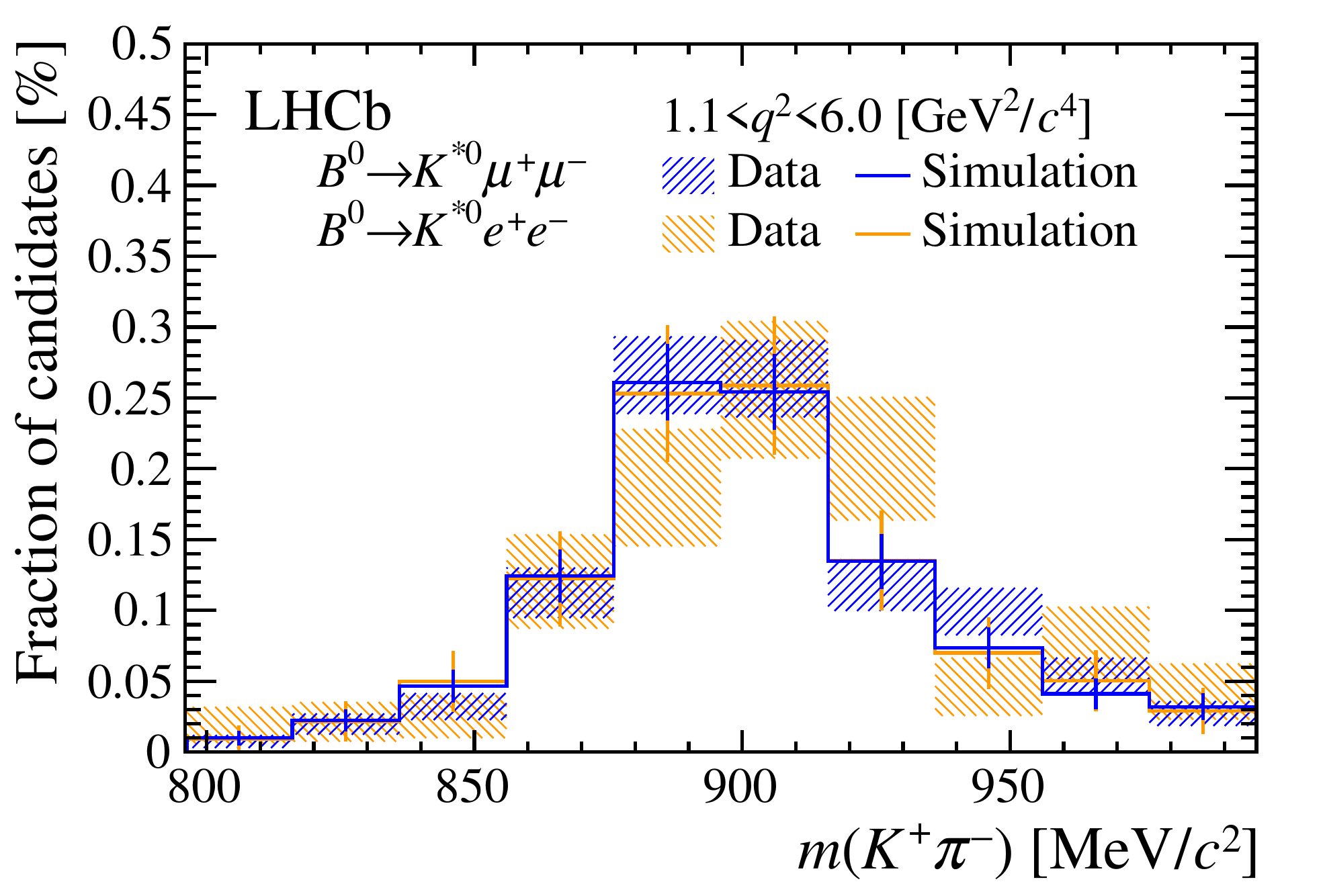}
\includegraphics[width=0.42\textwidth]{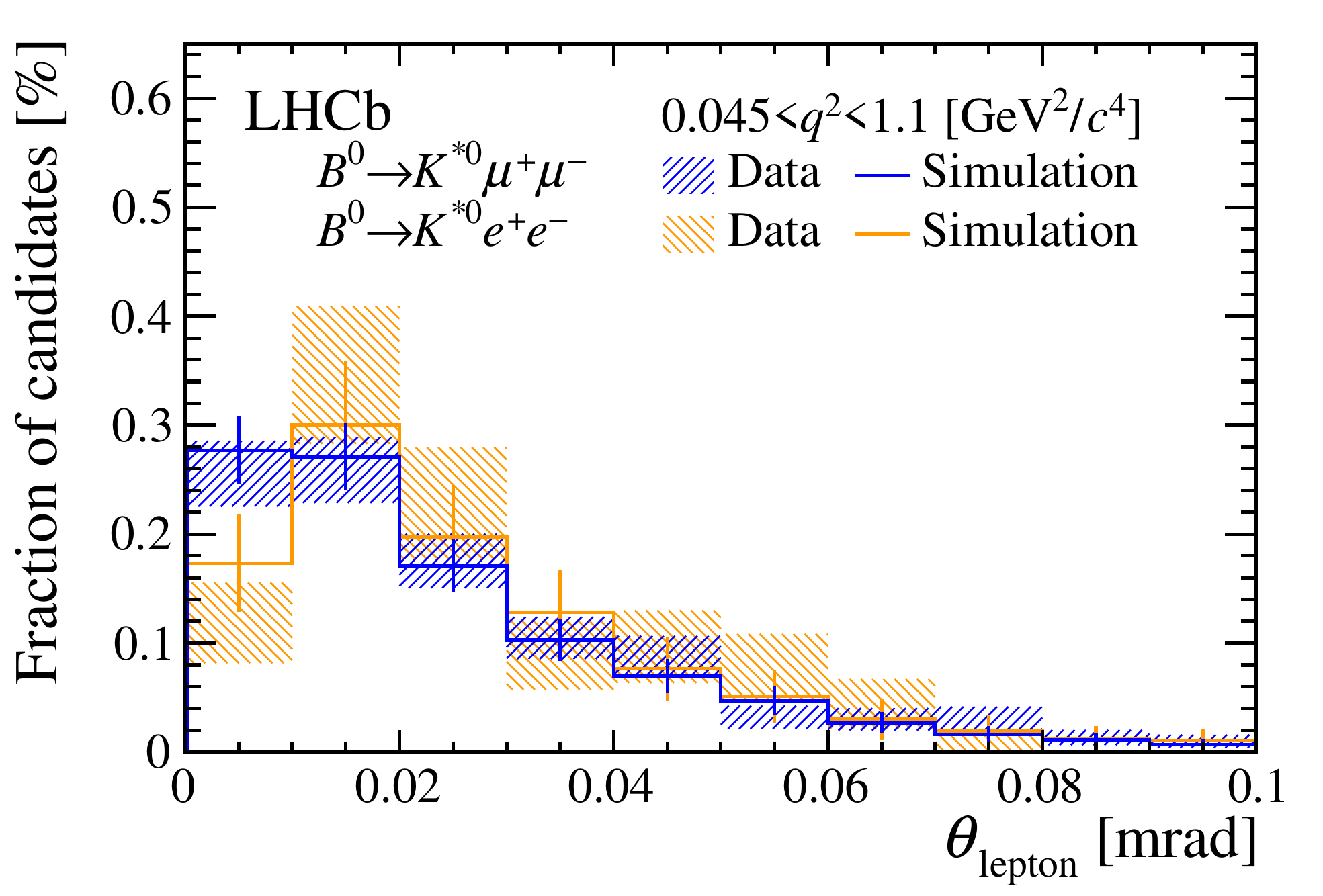}
\includegraphics[width=0.42\textwidth]{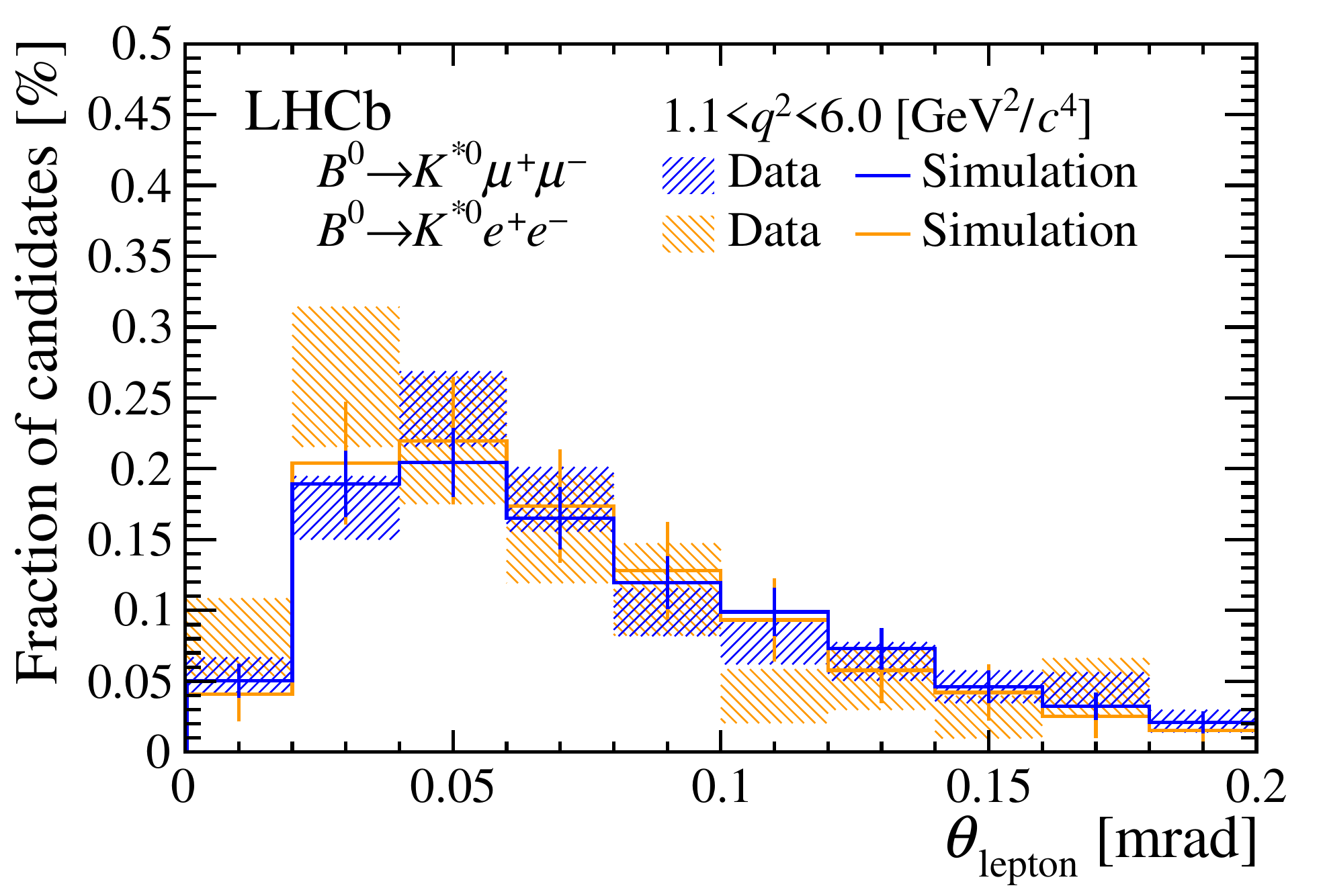}
\includegraphics[width=0.42\textwidth]{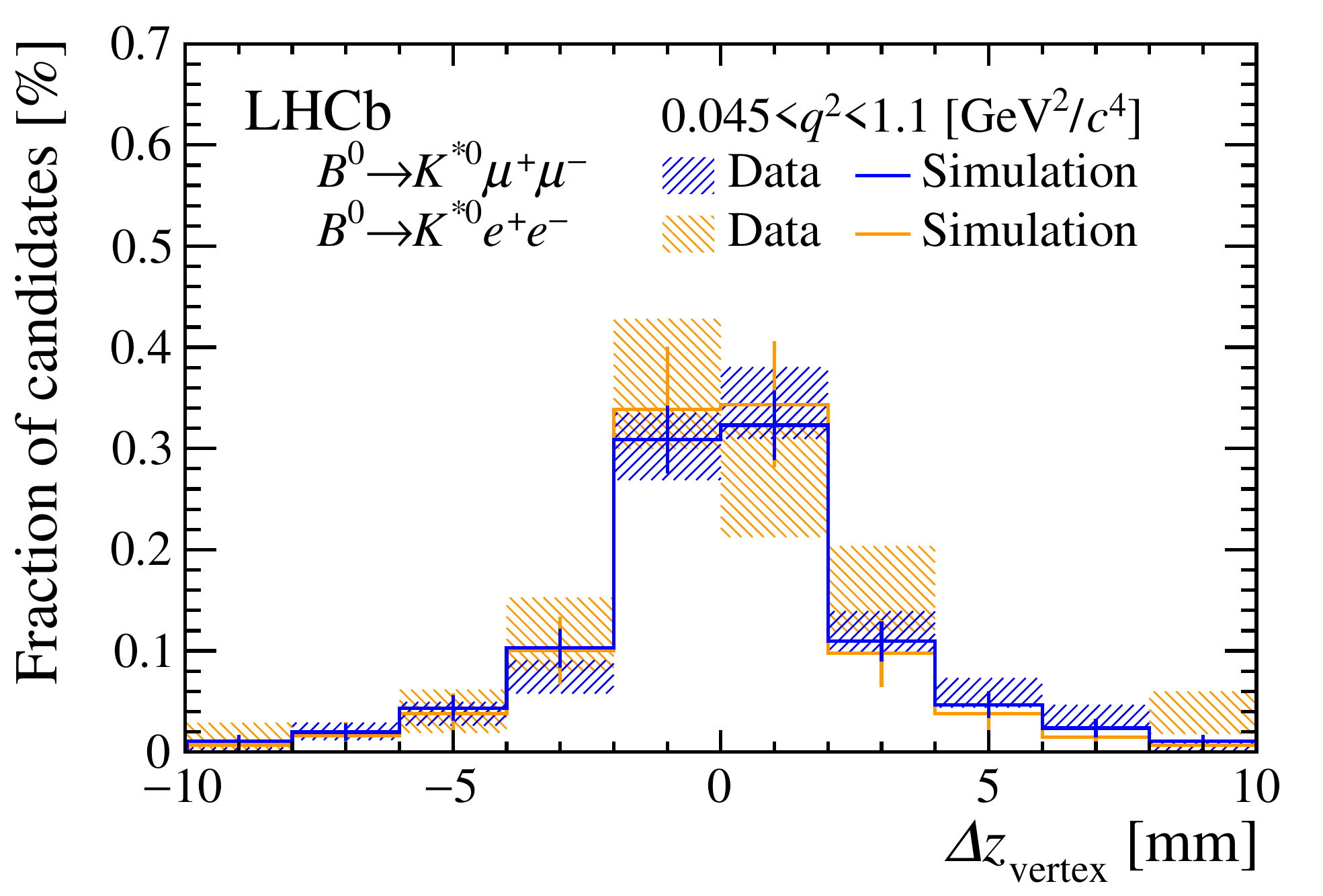}
\includegraphics[width=0.42\textwidth]{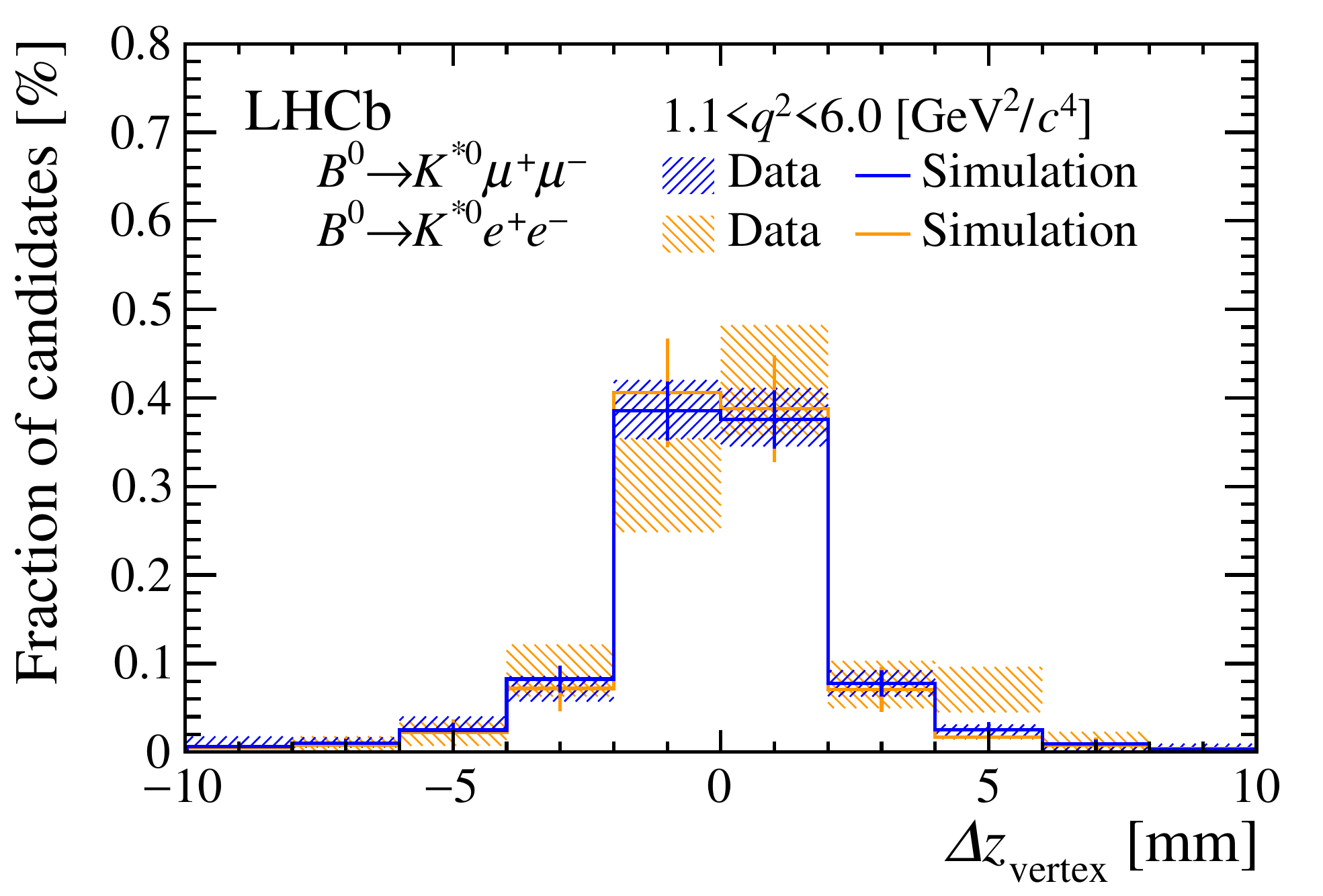}
\caption{(hatched) Background-subtracted distributions for (darker colour) \BdToKstmm and (lighter colour) \BdToKstee candidates, compared to (full line) simulation. From top to bottom: \qsq, \KPi invariant mass, \mKpi, opening angle between the two leptons, $\theta_{\textrm{lepton}}$, and projection along the beam axis of the distance between the \KPi and \ll vertices, $\Delta z_{\textrm{vertex}}$. The distributions are normalised to unity. The hatched areas correspond to the statistical uncertainties only. The data are not efficiency corrected.}
\label{fig:splot}
\end{figure}

The \sPlot\xspace technique~\cite{Pivk:2004ty}, where \mKpill is used as the discriminating variable, is adopted to subtract statistically the background from the \BdToKstll selected data, and test the agreement between muons and electrons, data and simulation, using several control quantities (see figure~\ref{fig:splot}):
the \qsq distributions show good agreement in both \qsq regions;
a clear \Kstarz peak is visible in the \KPi invariant mass distributions, and the muon and electron channels show good agreement;
while the distribution of the opening angle between the two leptons in the \cqsq region are very similar between the muon and electron channels, this is not the case at \lqsq due to the difference in lepton masses;
the distribution of the distance between the \KPi and \ll vertices shows that the pairs of hadrons and leptons consistently originate from the same decay vertex.

\section{Systematic uncertainties}
\label{sec:systematics}

Since \RKst is measured as a double ratio, many potential sources of systematic uncertainty cancel.
The remaining  systematics  and their effects on  \RKst  are summarised in table~\ref{tab:systematics} and are described below.

\begin{table}[t!]
\centering
\caption{Systematic uncertainties on the \RKst ratio for the three trigger categories separately (in percent). The total uncertainty is the sum in quadrature of all the contributions.}
\label{tab:systematics}
\renewcommand\arraystretch{1.4}
\begin{tabular}{c|ccc|ccc}
							& \multicolumn{6}{c}{$\Delta\RKst/\RKst$ [\%]} \\ \cline{2-7}
							& \multicolumn{3}{c|}{\lqsq}	& \multicolumn{3}{c}{\cqsq} \\ \hline
Trigger category		& \loe & \loh & \loi & \loe & \loh & \loi \\ \hline
Corrections to simulation	& 2.5 & 4.8 & 3.9 & 2.2 & 4.2 & 3.4\\
Trigger				& 0.1 & 1.2 & 0.1 & 0.2 & 0.8 & 0.2 \\
PID					& 0.2 & 0.4 & 0.3 & 0.2 & 1.0 & 0.5 \\
Kinematic selection		& 2.1 & 2.1 & 2.1 & 2.1 & 2.1 & 2.1 \\
Residual background	& -- & -- & -- & 5.0 & 5.0 & 5.0 \\
Mass fits				& 1.4 & 2.1 & 2.5 & 2.0 & 0.9 & 1.0 \\
Bin migration			& 1.0 & 1.0 & 1.0 & 1.6 & 1.6 & 1.6 \\
\RJPs ratio			& 1.6 & 1.4 & 1.7 & 0.7 & 2.1 & 0.7 \\
\hline
Total					& 4.0 & 6.1 & 5.5 & 6.4 & 7.5 & 6.7
\end{tabular}
\end{table}
\begin{description}

\item \textbf{Corrections to simulation}:
the uncertainty induced by the limited size of the simulated sample used to compute the efficiencies is considered;
an additional systematic uncertainty is determined using binned corrections instead of interpolated ones;
finally, since the data samples used to determine the corrections have a limited size, particularly for the electron hardware trigger, a systematic uncertainty is assessed with a bootstrapping technique~\cite{bootstrap}.

\item \textbf{Trigger efficiency}:
for the hardware triggers, the corrections to the simulation are determined using different control samples and
the change in the result is assigned as a systematic uncertainty;
for the software trigger, the corrections to the simulation do not show dependences on the kinematic of the decays, and therefore only the statistical uncertainty on the overall correction is considered as a systematic uncertainty. 

\item \textbf{Particle identification}:
the particle identification response is calibrated using data;
a systematic uncertainty due to the procedure and kinematic differences between these control samples and the signal modes is included;
the effects due to the identification of leptons and hadrons are considered;
however, discrepancies in the description of the latter are small and further cancel in the double ratio.

\item \textbf{Kinematic selection}:
a systematic uncertainty due to the choice of the mass fit range and to the two-dimensional requirement on \chisqfd and \mcorr is determined by comparing the efficiencies in simulation and background-subtracted samples of \BdToKstJPsmm or \BdToKstJPsee decays. 

\item \textbf{Residual background}:
background due to \BdToKstJPsee decays where one of the hadrons is misidentified as an electron and vice versa is studied;
using simulation that is tuned to data (see section~\ref{sec:weights}) this contribution is estimated to be small;
however, a few candidates with one electron of the dilepton pair having a low probability to be genuine are observed in background subtracted data;
a systematic uncertainty is assigned based on the distribution of the PID information of these candidates.

\item \textbf{Mass fit}:
the systematic uncertainty due to the parameterisation of the signal invariant mass distributions is found to be negligible for the muon channel;
for the electron channel, the signal PDF is changed from the sum of a CB and a Gaussian function to the sum of two CB functions, where the mean parameter is shared and, additionally, the mass shift and the width scale factors are constrained using the \BdToKstGee decay mode instead of \BdToKstJPsee;
the relative fractions of the three bremsstrahlung categories are measured in data using \BdToKstJPsee and the observed differences with respect to simulation are used in the mass fit (see figure~\ref{fig:brem});
for the backgrounds, a component that describes candidates where the hadron identities are swapped is added both to the muon and electron \BdToKstJPsll modes, and constrained to the expected values observed in simulation;
the kernel of the nonparametric models is also varied, as well as the mixture of the \Kone and \Ktwo components that is constrained using data~\cite{LHCb-PAPER-2014-030};
the contributions to the systematic uncertainty from these sources are evaluated using pseudoexperiments that are generated with modified parameters and fitted with the PDFs used to fit the data.

\item \textbf{Bin migration}:
for the electron channel, the degraded \qsq resolution due to bremsstrahlung emission causes a nonnegligible fraction of signal candidates to migrate in and out of the given \qsq bin;
the effect is included in the efficiency determination, but introduces a small dependence on the shape of the differential branching fraction that no longer perfectly cancels in the ratio to the muon channel;
pseudoexperiments are generated, where the parameters modelling the $d\Gamma(\BdToKstee)/d\qsq$ distribution are varied within their uncertainties~\cite{flavio};
the maximum spread of the variation in \RKst is taken as a systematic uncertainty;
furthermore, the \qsq resolution is smeared for differences between data and simulation that are observed in the resonant mode.

\item \textbf{$\boldsymbol{\RJPs}$ ratio}:
the ratio of the efficiency-corrected yield of the resonant modes (see section~\ref{sec:crosschecks}) is expected to be unity to a very high precision;
deviations from unity are therefore considered to be a sign of residual imperfections in the evaluation of the efficiencies;
the \RJPs ratio is studied as a function of various event and kinematic properties of the decay products, and the observed residual deviations from unity are used to assign a systematic uncertainty on \RKst.

\end{description}

For the \RKst measurement, all the uncertainties are treated as uncorrelated among the trigger categories, except for those related to particle identification, to the kinematic selection criteria, to the residual background, to the fit to the invariant mass and to bin migration.
 
%%%%%%%%%%%%%%%%%%%%%%%%%%%%%%%%%%%%
% !TEX root = main.tex
%%%%%%%%%%%%%%%%%%%%%%%%%%%%%%%%%%%%

\section{Results}
\label{sec:results}

\begin{figure}[t!]
\centering
\includegraphics[width=0.49\textwidth]{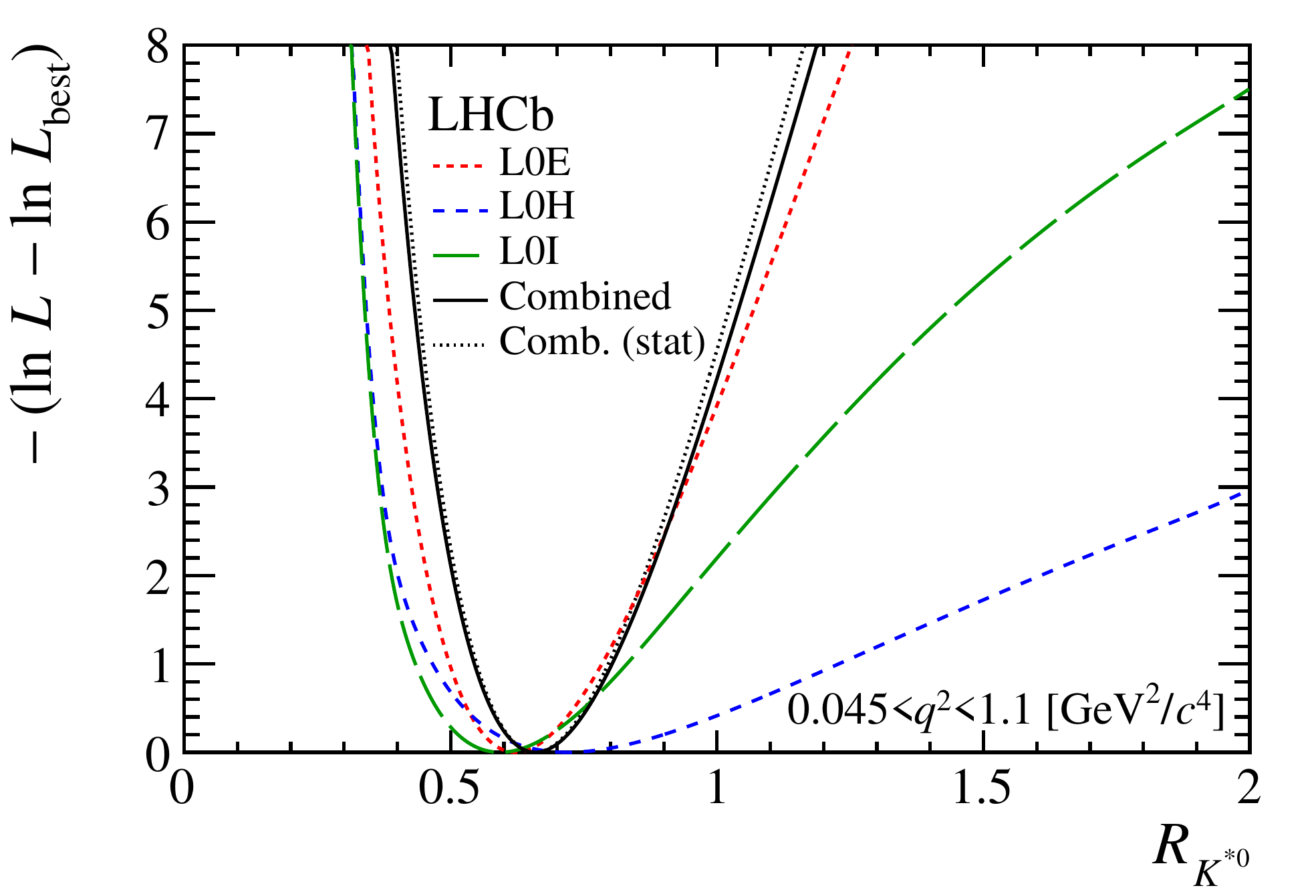}
\includegraphics[width=0.49\textwidth]{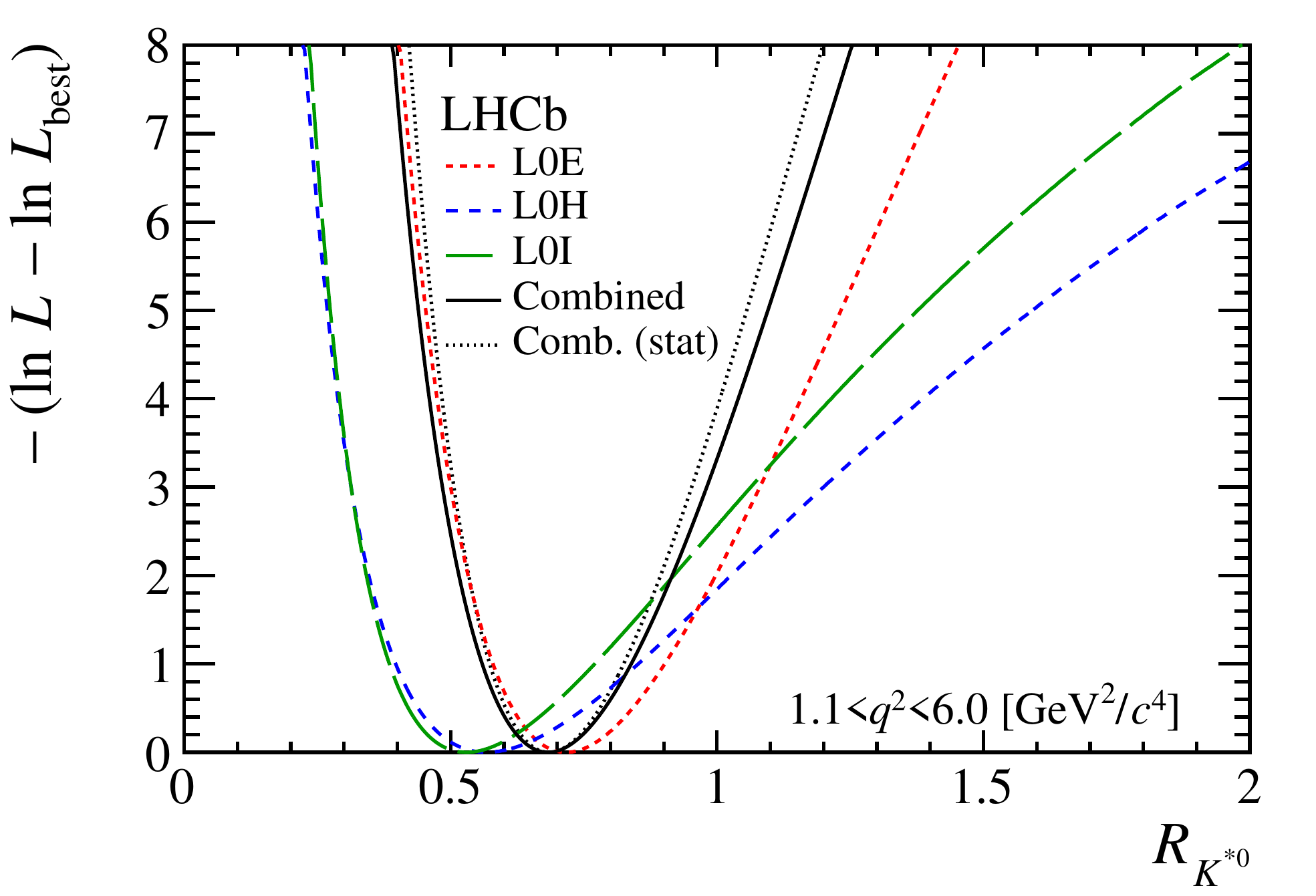}
\caption{Distributions of the \RKst delta log-likelihood for the three trigger categories separately and combined.}
\label{fig:LL}
\end{figure}

\begin{figure}[t!]
\centering
\includegraphics[width=0.49\textwidth]{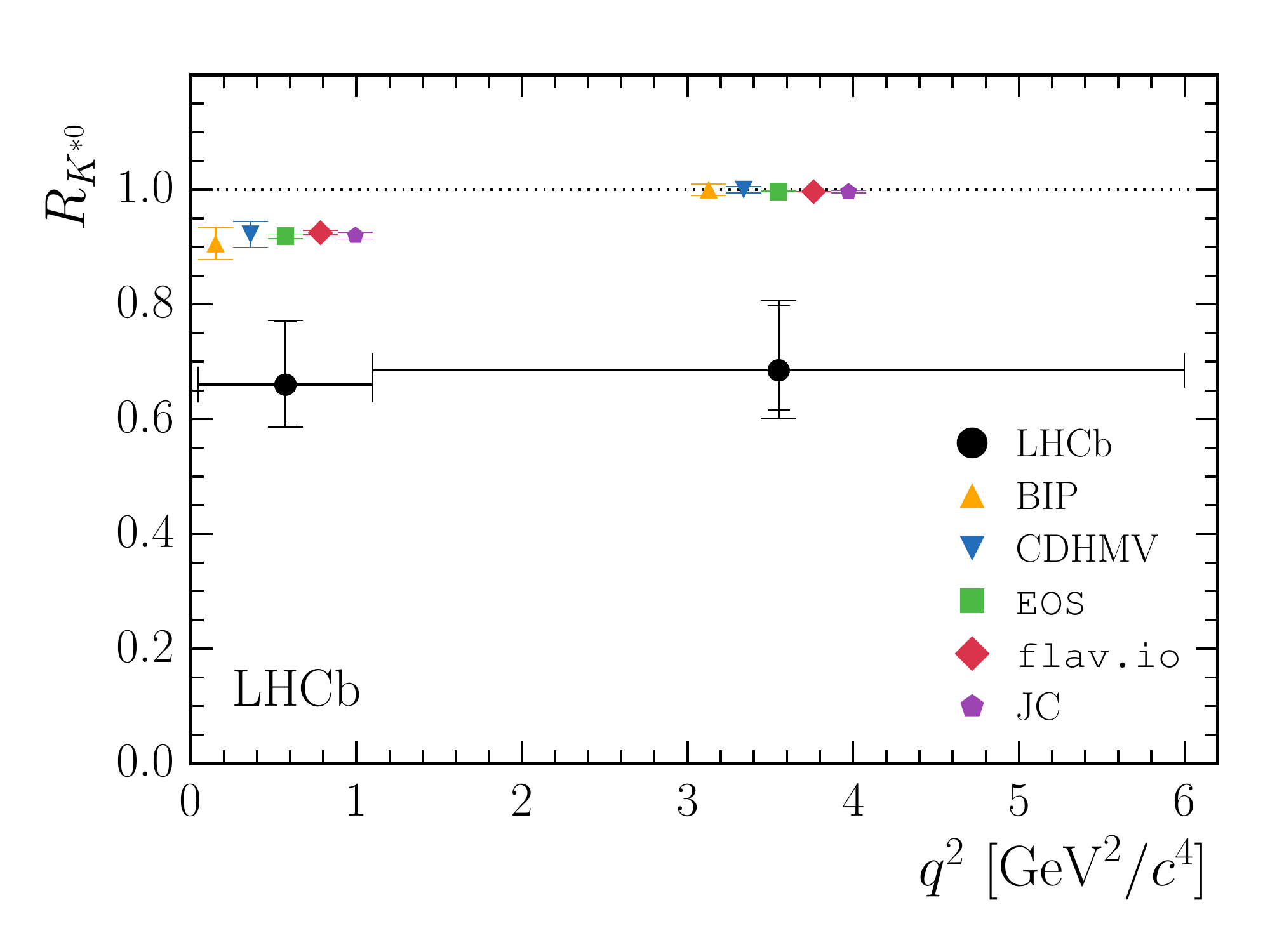}
\includegraphics[width=0.49\textwidth]{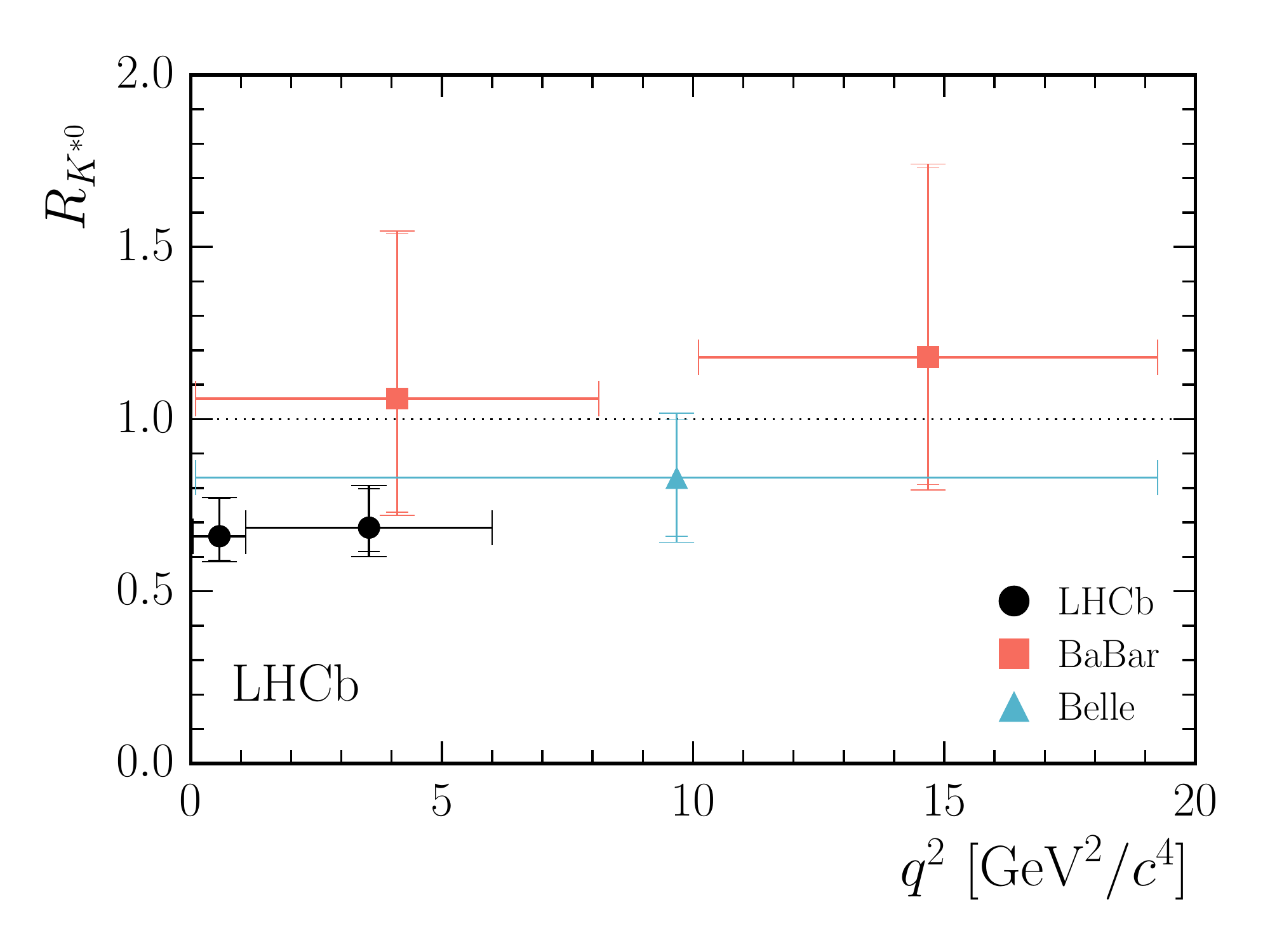}
\caption{(left) Comparison of the \lhcb \RKst measurements with the SM theoretical predictions: BIP~\cite{Bordone:2016gaq} CDHMV~\cite{Descotes-Genon:2015uva,Capdevila:2016ivx,Capdevila:2017ert}, \texttt{EOS}~\cite{Serra:2016ivr,EOS-Web,*EOS}, \texttt{flav.io}~\cite{Straub:2015ica,Altmannshofer:2017fio,flavio} and JC~\cite{Jager:2014rwa}. The predictions are displaced horizontally for presentation. (right) Comparison of the \lhcb \RKst measurements with previous experimental results from the \B factories~\cite{Lees:2012tva,Wei:2009zv}. In the case of the \B factories the specific vetoes for charmonium resonances are not represented.}
\label{fig:results}
\end{figure}

The determination of \RKst exploits the log-likelihoods resulting from the fits to the invariant mass distributions of the nonresonant and resonant channels in each trigger category and \qsq region.
Each log-likelihood is used to construct the PDF of the true number of decays, which is used as a prior to obtain the PDF of \RKst.
The true number of decays is assumed to have a uniform prior.
The three electron trigger categories are combined by summing the corresponding log-likelihoods.
Uncorrelated systematic uncertainties are accounted for by convolving the yield PDFs with a Gaussian distribution of appropriate width.
Correlated systematic uncertainties are treated by convolving the \RKst PDF with a Gaussian distribution. 
The one, two and three standard deviation intervals are determined as the ranges that include 68.3\%, 95.4\% and 99.7\% of the PDF.
In each \qsq region, the measured values of \RKst are found to be in good agreement among the three electron trigger categories (see figure~\ref{fig:LL}).
The results are given in table~\ref{tab:results} and presented in figure~\ref{fig:results}, where they are compared both to the SM predictions (see table~\ref{tab:predictions}) and to previous measurements from the \B factories~\cite{Lees:2012tva,Wei:2009zv}.

The combined \RKst PDF is used to determine the compatibility with the SM expectations.
The $p$-value, calculated by integrating the PDF above the expected value, is translated into a number of standard deviations. 
The compatibility with the SM expectations~\cite{Bordone:2016gaq,Descotes-Genon:2015uva,Capdevila:2016ivx,Capdevila:2017ert,Serra:2016ivr,EOS-Web,*EOS,Straub:2015ica,Altmannshofer:2017fio,flavio,Jager:2014rwa} is determined to be 2.1--2.3 and 2.4--2.5 standard deviations, for the low-\qsq and the  central-\qsq  regions, respectively, depending on the theory prediction used.

\begin{table}[t!]
\centering
\caption{Measured \RKst ratios in the two \qsq regions. The first uncertainties are statistical and the second are systematic. About 50\% of the systematic uncertainty is correlated between the two \qsq bins. The 95.4\% and 99.7\% confidence level (CL) intervals include both the statistical and systematic uncertainties.}
\label{tab:results}
\renewcommand\arraystretch{1.4}
\begin{tabular}{c|c|c}
			& \lqsq							& \cqsq \\ \hline 
\RKst		& $0.66~^{+~0.11}_{-~0.07} \pm 0.03$	& $0.69~^{+~0.11}_{-~0.07} \pm 0.05$ \\ \hline
95.4\% CL		& $[0.52,0.89]$						& $[0.53,0.94]$ \\
99.7\% CL		& $[0.45,1.04]$						& $[0.46,1.10]$ \\
\end{tabular}
\end{table}

\section{Conclusions}
\label{sec:conclusions}

This paper reports a test of lepton universality performed by measuring the ratio of the branching fractions of the decays \BdToKstmm and \BdToKstee.
The \Kstarz meson is reconstructed in the final state \KPi, which is required to have an invariant mass within 100\mevcc of the known $\Kstar(892)^{0}$ mass. 
Data corresponding to an integrated luminosity of 3\invfb of \pp collisions, recorded by the \lhcb experiment during 2011 and 2012, are used.
The \RKst ratio is measured in two regions of the dilepton invariant mass squared to be
\begin{eqnarray*}
\RKst = 
\begin{cases}
0.66~^{+~0.11}_{-~0.07}\stat  \pm 0.03\syst	& \textrm{for } 0.045 < \qsq < 1.1~\gevgevcccc \, , \\
0.69~^{+~0.11}_{-~0.07}\stat \pm 0.05\syst	& \textrm{for } 1.1\phantom{00} < \qsq < 6.0~\gevgevcccc \, .
\end{cases}
\end{eqnarray*}
The corresponding 95.4\% confidence level intervals are $[0.52, 0.89]$ and $[0.53, 0.94]$. 
The results, which represent the most precise measurements of \RKst to date, are compatible with the SM expectations~\cite{Bordone:2016gaq,Descotes-Genon:2015uva,Capdevila:2016ivx,Capdevila:2017ert,Serra:2016ivr,EOS-Web,*EOS,Straub:2015ica,Altmannshofer:2017fio,flavio,Jager:2014rwa} at 2.1--2.3 standard deviations for the \lqsq region and 2.4--2.5 standard deviations for the central-\qsq region, depending on the theoretical prediction used.

Model-independent fits to the ensemble of FCNC data that allow for NP contributions~\cite{Descotes-Genon:2015uva,Capdevila:2016ivx,Capdevila:2017ert,Serra:2016ivr,EOS-Web,*EOS,Straub:2015ica,Altmannshofer:2017fio,flavio,Jager:2014rwa} lead to predictions for \RKst in the \cqsq region that are similar to the value observed; smaller deviations are expected at \lqsq.
The larger data set currently being accumulated by the \lhcb collaboration will allow for more precise tests of these predictions.

\section*{Acknowledgements}

\noindent We would like to thank F.~Le~Diberder for many interesting and helpful discussions on statistics.
We express our gratitude to our colleagues in the CERN
accelerator departments for the excellent performance of the LHC. We
thank the technical and administrative staff at the LHCb
institutes. We acknowledge support from CERN and from the national
agencies: CAPES, CNPq, FAPERJ and FINEP (Brazil); MOST and NSFC (China);
CNRS/IN2P3 (France); BMBF, DFG and MPG (Germany); INFN (Italy);
NWO (The Netherlands); MNiSW and NCN (Poland); MEN/IFA (Romania);
MinES and FASO (Russia); MinECo (Spain); SNSF and SER (Switzerland);
NASU (Ukraine); STFC (United Kingdom); NSF (USA).
We acknowledge the computing resources that are provided by CERN, IN2P3 (France), KIT and DESY (Germany), INFN (Italy), SURF (The Netherlands), PIC (Spain), GridPP (United Kingdom), RRCKI and Yandex LLC (Russia), CSCS (Switzerland), IFIN-HH (Romania), CBPF (Brazil), PL-GRID (Poland) and OSC (USA). We are indebted to the communities behind the multiple open
source software packages on which we depend.
Individual groups or members have received support from AvH Foundation (Germany),
EPLANET, Marie Sk\l{}odowska-Curie Actions and ERC (European Union),
Conseil G\'{e}n\'{e}ral de Haute-Savoie, Labex ENIGMASS and OCEVU,
R\'{e}gion Auvergne (France), RFBR and Yandex LLC (Russia), GVA, XuntaGal and GENCAT (Spain), Herchel Smith Fund, The Royal Society, Royal Commission for the Exhibition of 1851 and the Leverhulme Trust (United Kingdom). 

\addcontentsline{toc}{section}{References}
\setboolean{inbibliography}{true}
\bibliographystyle{LHCb}
\bibliography{main,LHCb-PAPER,LHCb-CONF,LHCb-DP,LHCb-TDR}

\clearpage
\centerline{\large\bf LHCb collaboration}
\begin{flushleft}
\small
R.~Aaij$^{40}$,
B.~Adeva$^{39}$,
M.~Adinolfi$^{48}$,
Z.~Ajaltouni$^{5}$,
S.~Akar$^{59}$,
J.~Albrecht$^{10}$,
F.~Alessio$^{40}$,
M.~Alexander$^{53}$,
S.~Ali$^{43}$,
G.~Alkhazov$^{31}$
% ...

{\footnotesize \it
% ...
$ ^{5}$Clermont Universit{\'e}, Universit{\'e} Blaise Pascal, CNRS/IN2P3, LPC, Clermont-Ferrand, France\\
% ...
$ ^{10}$Fakult{\"a}t Physik, Technische Universit{\"a}t Dortmund, Dortmund, Germany\\
% ...
$ ^{31}$Petersburg Nuclear Physics Institute (PNPI), Gatchina, Russia\\
% ...
$ ^{39}$Universidad de Santiago de Compostela, Santiago de Compostela, Spain\\
$ ^{40}$European Organization for Nuclear Research (CERN), Geneva, Switzerland\\
% ...
$ ^{43}$Nikhef National Institute for Subatomic Physics, Amsterdam, The Netherlands\\
% ...
$ ^{48}$H.H. Wills Physics Laboratory, University of Bristol, Bristol, United Kingdom\\
% ...
$ ^{53}$School of Physics and Astronomy, University of Glasgow, Glasgow, United Kingdom\\
% ...
$ ^{59}$University of Cincinnati, Cincinnati, OH, United States\\
% ...
}
\end{flushleft}
%%%%%%%%%%%%%%%%%%%%%%%%%
\end{document}